\documentclass[aps,prl,twocolumn,groupedaddress,showpacs,superscriptaddress]{revtex4-2}
\usepackage{graphicx,amsmath,amsthm,amssymb,color}
\usepackage[colorlinks=true,citecolor=blue,linkcolor=blue,urlcolor=magenta]{hyperref}
\usepackage{verbatim}
\usepackage{physics}

\begin{document}

    \title{Robust Quantum Gates against Correlated Noise in Integrated Quantum Chips}
    
    \author{Kangyuan Yi}
    \thanks{K.-Y. Y. and Y.-J. H. contributed equally to this work.}
    \affiliation{Department of Physics, Southern University of Science and Technology, Shenzhen 518055, China}
    \author{Yong-Ju Hai}
    \thanks{K.-Y. Y. and Y.-J. H. contributed equally to this work.}
    \affiliation{Shenzhen Institute for Quantum Science and Engineering, Southern University of Science and Technology, Shenzhen 518055, China}
    \affiliation{International Quantum Academy (SIQA), Shenzhen 518048, China}
    \author{Kai Luo}
       \affiliation{Department of Physics, Southern University of Science and Technology, Shenzhen 518055, China}
	\affiliation{Shenzhen Institute for Quantum Science and Engineering, Southern University of Science and Technology, Shenzhen 518055, China}
     \author{Ji Chu}
      \affiliation{Shenzhen Institute for Quantum Science and Engineering, Southern University of Science and Technology, Shenzhen 518055, China}
    \affiliation{International Quantum Academy (SIQA), Shenzhen 518048, China}
    \author{Libo Zhang}
    \affiliation{Shenzhen Institute for Quantum Science and Engineering, Southern University of Science and Technology, Shenzhen 518055, China}
    \affiliation{International Quantum Academy (SIQA), Shenzhen 518048, China}
    \author{Yuxuan Zhou}
    \affiliation{Department of Physics, Southern University of Science and Technology, Shenzhen 518055, China}
	\affiliation{Shenzhen Institute for Quantum Science and Engineering, Southern University of Science and Technology, Shenzhen 518055, China}
    \author{Yao Song}
    \affiliation{Shenzhen Institute for Quantum Science and Engineering, Southern University of Science and Technology, Shenzhen 518055, China}
    \affiliation{International Quantum Academy (SIQA), Shenzhen 518048, China}
    \author{Song Liu}
    \affiliation{Shenzhen Institute for Quantum Science and Engineering, Southern University of Science and Technology, Shenzhen 518055, China}
    \affiliation{International Quantum Academy (SIQA), Shenzhen 518048, China}
    \affiliation{Guangdong Provincial Key Laboratory of Quantum Science and Engineering, Southern University of Science and Technology, Shenzhen, 518055, China}
    \author{Tongxing Yan}
	\email{yantx@sustech.edu.cn}
    \affiliation{Shenzhen Institute for Quantum Science and Engineering, Southern University of Science and Technology, Shenzhen 518055, China}
    \affiliation{International Quantum Academy (SIQA), Shenzhen 518048, China}
    \affiliation{Guangdong Provincial Key Laboratory of Quantum Science and Engineering, Southern University of Science and Technology, Shenzhen, 518055, China}
 
    \author{Xiu-Hao Deng}
    \email{dengxh@sustech.edu.cn}
    \affiliation{Shenzhen Institute for Quantum Science and Engineering, Southern University of Science and Technology, Shenzhen 518055, China}
    \affiliation{International Quantum Academy (SIQA), Shenzhen 518048, China}
    \affiliation{Guangdong Provincial Key Laboratory of Quantum Science and Engineering, Southern University of Science and Technology, Shenzhen, 518055, China}
    
    \author{Yuanzhen Chen}
    \email{chenyz@sustech.edu.cn}
	\affiliation{Department of Physics, Southern University of Science and Technology, Shenzhen 518055, China}
	\affiliation{Shenzhen Institute for Quantum Science and Engineering, Southern University of Science and Technology, Shenzhen 518055, China}
    \affiliation{Guangdong Provincial Key Laboratory of Quantum Science and Engineering, Southern University of Science and Technology, Shenzhen, 518055, China}
    \author{Dapeng Yu}
   \affiliation{Department of Physics, Southern University of Science and Technology, Shenzhen 518055, China}
    \affiliation{Shenzhen Institute for Quantum Science and Engineering, Southern University of Science and Technology, Shenzhen 518055, China}
    \affiliation{International Quantum Academy (SIQA), Shenzhen 518048, China}
    \affiliation{Guangdong Provincial Key Laboratory of Quantum Science and Engineering, Southern University of Science and Technology, Shenzhen, 518055, China}
	
	\date{\today }

    \begin{abstract}
    As quantum circuits become more integrated and complex, additional error sources that were previously insignificant start to emerge. Consequently, the fidelity of quantum gates benchmarked under pristine conditions falls short of predicting their performance in realistic circuits. To overcome this problem, we must improve their robustness against pertinent error models besides isolated fidelity. Here we report the experimental realization of robust quantum gates in superconducting quantum circuits based on a geometric framework for diagnosing and correcting various gate errors. Using quantum process tomography and randomized benchmarking, we demonstrate robust single-qubit gates against quasi-static noise and spatially correlated noise in a broad range of strengths, which are common sources of coherent errors in large-scale quantum circuits. We also apply our method to non-static noises and to realize robust two-qubit gates. Our work provides a versatile toolbox for achieving noise-resilient complex quantum circuits. 
    \end{abstract}
	
	\maketitle
	
    Quantum logic gates are typically benchmarked in isolation under pristine conditions to obtain high fidelities surpassing fault-tolerance thresholds of quantum error correction (QEC) codes~\cite{cheng2023noisy,bharti2022noisy,alexeev2021quantum}. However, when deployed in large-scale quantum circuits, additional noise channels emerge, leading to errors that are absent or negligible in the isolated gate setting~\cite{monroe2021programmable,bluvstein2022quantum,arute2019quantum,bravyi2021mitigating}. These noises arise from effects like crosstalk, control imperfection, and correlated noise \cite{mundada2019suppression,sheldon2016procedure,gambetta2012characterizationa,reed2012realization,bialczak2007flux,gustavsson2011noise,rower2023evolution}, leading to complex errors that are difficult to benchmark in isolation. Consequently, the gate fidelity measured under well-controlled conditions fails to faithfully predict performance in real circuits. To overcome this challenge, we need to rigorously evaluate the gate robustness against pertinent error models beyond isolated fidelity \cite{harrow2003robustness}.

    Robust gates exhibit built-in noise resilience through careful pulse shaping, making them well-suited for constructing complex quantum algorithms \cite{dridi2020optimal, liu2021superrobust,le2023scalablea}. In particular, they help correct coherent errors of significant spatiotemporal correlations, which are known to pose great challenges for QEC~\cite{harper2023learning, iverson2020coherence, kueng2016comparing}. Recently, a geometric technique has been developed to design smooth and short-duration robust control pulses (RCPs) that suppress errors from general noise processes in multiple directions~\cite{zeng2019geometric, dong2021doubly, buterakos2021geometrical, deng2021correcting, hai2022universal}. The RCPs are constructed by mapping the noisy quantum evolution onto error curves in a parameter space. The topology of these curves, such as their closeness, directly determines the gate robustness, while the local geometric properties like curvature and torsion are linked to parameters of the control Hamiltonian. This framework provides both an intuitive picture of noisy quantum evolution and a systematic tool for RCP optimization.

    In this work, we experimentally demonstrate robust quantum gates based on the above RCPs using superconducting quantum circuits. After a concise introduction to the design principles, we report experimental results on quantum gates that are much more fault-tolerant against coherent errors resulting from generic noises with components in multiple directions than conventional dynamical gates. More importantly, we find that our gates significantly suppress the pernicious buildup of coherent errors in long circuits subjected to temporally correlated noise, leading to substantially enhanced overall circuit performance and worst-case fidelity, which may benefit fault-tolerant QEC.
    
    \begin{figure}[t]
       \centering
       \includegraphics[width =0.48\textwidth]{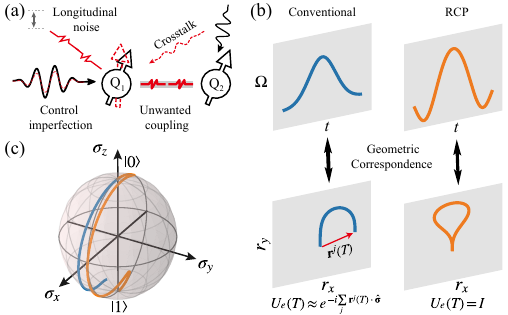}
       \caption{(\textbf{a}) Schematics of the realistic noisy environment of qubits. (\textbf{b}) Upper panel: Schematics of the control signal envelopes, $\Omega(t)$, for a Gaussian pulse (blue) and an RCP (orange). Lower panel: Error curves of the two pulses derived in our geometric framework, in the presence of a generic noise as given in the main text. The red vector $\mathbf{r}^j(T)$ indicates a finite susceptibility to noise, corresponding to a compromised robustness. Conversely, a closed error curve in the RCP case (orange) yields a robust gate to the first order. (\textbf{c}) Dynamics of an $X^{\pi}$ gate for an initial state of $\vert 0\rangle$ using Gaussian pulse (blue) and RCP (orange) in the presence of a static frequency detuning noise.}
       \label{Fig1_protocol}
    \end{figure}
    
    We start with the Hamiltonian of a qubit driven by a control field $H_{c}(t)$ and subjected to a generic noise $V(t)$: $H(t) = H_{0} + H_{c}(t) + V(t)$, where $H_{0}$ is the free-qubit Hamiltonian. The noiseless evolution is $U_{0}(t)=\mathcal{T}\exp \{-i\int_{0}^{t}(H_{0}+H_{c}(\tau))d\tau \}$ ($\mathcal{T}$ stands for time ordering). We define an error unitary as $U_e(t)=\mathcal{T} \exp \{-i \int_0^t d \tau H_I(\tau)\}$ where $H_I = U_0^{\dagger} V U_0$ is the noise in the interaction frame determined by $U_0$. Since the total evolution unitary can be decomposed as $U = U_0 U_e$ (see SI~\cite{SI}), it then follows that to realize robust quantum gates, one simply engineers $H_{c}(t)$ to obtain a target gate with a duration of $T$, $U_0(T) = U_{\text{target}}$ while eliminating the impact of noise at the end of evolution, i.e. $U_e(T) = I$.
       
    To proceed, we write the noise term as $V(t) = \sum_{j,k} \epsilon^j_k v^j_k(t)\sigma_{k}$ where $k = x,y,z$ and $\epsilon^j_k$ is a time-independent amplitude of the $k$-component for the $j$-th noise source and $v^j_k$ is its possibly time-dependent profile. $\sigma_k$ are the Pauli matrices. Hereafter, we reserve the upper index $j$ for numbering different noise sources. Some of the possible origins of the noises are illustrated in Fig.~\ref{Fig1_protocol}(a). The error unitary $U_e(t)$ can be calculated using a perturbative expansion (see SI~\cite{SI}). Up to the leading order, one has $U_e(t) \approx  e^{-i \sum_j \Phi^j(t)}$, where $\Phi^j(t)$ is a matrix associated with the $j$-th noise source and can be expressed as $\Phi^j(t)= \mathbf{r}^j(t) \cdot \hat{\boldsymbol{\sigma }}$. Here, $\mathbf{r}^j(t)$ is a vector that traces out a three-dimensional curve as $t$ evolves. This curve describes the accumulated error of the $j$-th noise on the evolution unitary and is thus named as an \textit{error curve}. The geometric correspondence between the noisy quantum dynamics and the error curves enables a straightforward and intuitive approach to designing robust quantum control. For example, the \textit{error distance} $R^j(t) = ||\mathbf{r}^j(t)||$ measures the susceptibility of the quantum dynamics to the $j$-th noise source and is thus a natural metric for characterizing the robustness of the corresponding quantum operation. Essentially, a robust control requires the error distances to vanish at the end of gate operation, i.e., $R^j(T) = 0$. With this condition met, one has $\Phi^j(T)=0$ and thus $U_e(T) = I$. We illustrate this robust control framework in Fig.~\ref{Fig1_protocol}(b)(c). Further details of the construction of robust control pulses can be found in the SI~\cite{SI} and Ref.~\cite{hai2022universal}. 

    \begin{figure}[!t]
		\centering
		\includegraphics[width =0.48\textwidth]{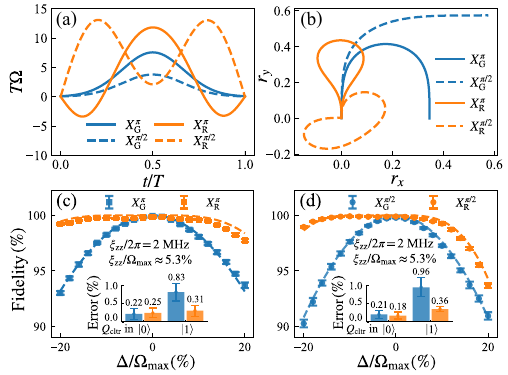}
		\caption{Robustness of representative single-qubit gates exposed to a static noise in the $z$ direction. (\textbf{a}) Pulse envelopes of two Gaussian (G) gates $X^{\pi}_G,X^{\pi/2}_G$ and two robust (R) gates $X^{\pi}_R,X^{\pi/2}_R$. For comparison, the horizontal axis is set to normalized time, with the actual gate time being $T = 32,16,50,55$ ns, respectively. The vertical axis represents a dimensionless quantity of $T\Omega(t)$ where $\Omega(t)$ is the actual time-dependent envelope. The maximal amplitude of the pulses is $\Omega_{\mathrm{max}}/2\pi = 37.5 $ MHz for all four gates. (\textbf{b}) The error curves of the four pulses in (\textbf{a}) for the frequency noise ($\mathbf{r}^{\Delta}$). (\textbf{c-d}) Main panel: gate fidelity characterized by QPT for the four gates. The experimental data (symbols) fit the numerical simulations (dashed lines) well. Error bars are the standard deviation. Note that all the gates here are implemented with DRAG \cite{motzoi2009simple,gambetta2011analytic} to suppress the leakage. More details about the simulation and experiment are given in the SI~\cite{SI}. Insets: gate error determined by QPT for an $X^{\pi}$ and an $X^{\pi/2}$ gates using the RCP and Gaussian pulses in the presence of a $ZZ$ interaction.}\label{Fig2}
    \end{figure}
    
    We have performed experiments on different superconducting quantum processors and obtained similar results. The data presented here were all acquired on a processor that consists of 8 transmon~\cite{koch2007chargeinsensitive, schreier2008suppressing} qubits arranged in a circle. Each qubit has a fixed frequency and is connected to the two nearest qubits via couplers that are also transmons but with tunable frequencies. More detailed information about this processor can be found in SI~\cite{SI}.
    
    We first demonstrate robust single-qubit quantum gates against quasi-static noise in the $z$ direction. For a transmon qubit, an equivalent quasi-static noise can be readily generated by purposely driving the qubit at a detuned frequency, leading to an effective Hamiltonian $H = 1/2\Omega(t) \sigma_x + 1/2 \Delta \sigma_z$, where $\Omega(t)$ and $\Delta$ are pulse amplitude and frequency detuning, respectively. Figure~\ref{Fig2} shows a representative data set where the performance of four different quantum gates are compared using the standard quantum process tomography (QPT)~\cite{nielsen_chuang_2010,obrien2004quantum,chow2009randomized}, two using conventional Gaussian drives and the other two using RCPs. 

    We also benchmark gate performance in the presence of a $ZZ$-type of crosstalk between neighboring qubits, which is a typical source for spatially-correlated errors in multi-qubit circuits. For this purpose, two qubits and a tunable coupler are used. The coupler is tuned to induce a variable $ZZ$ interaction between the two qubits. Consequently, the target qubit to be benchmarked feels a disturbance that depends on the state of the control qubit. The insets of Fig.\ref{Fig2}(c) and (d) plot the errors of an $X^{\pi}$ and an $X^{\pi/2}$ gates using the RCP and Gaussian pulses for a $ZZ$ interaction of 2 MHz, which corresponds to a relative noise magnitude of 5.3$\%$ ($\xi_{\mathrm{ZZ}}$/$\Omega_{\mathrm{max}}$). As the control qubit is set to different states, the gate performance of the target qubit is much more robust in the case of RCP pulses. A similar robust profile against the $ZZ$ interaction using $X^{\pi}_R$ gate was also observed in the fixed-frequency transmon system with always-on couplings~\cite{watanabe2024zz}.

    \begin{figure}[t]
		\centering
		\includegraphics[width =0.48\textwidth]{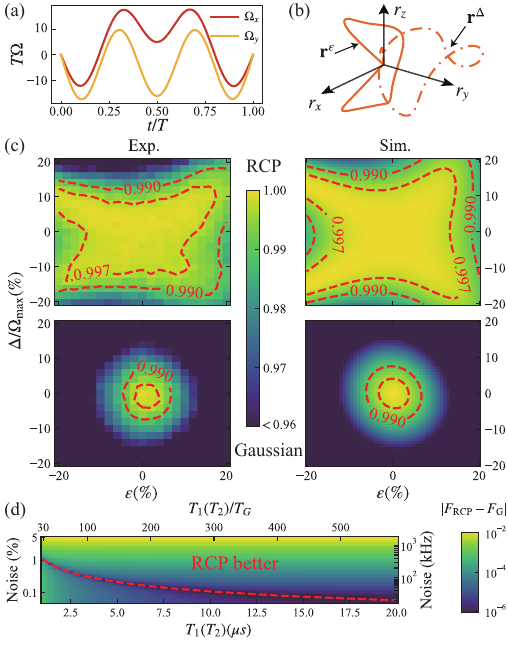}
		\caption{Robust gate against both control imperfection $\epsilon$ and frequency detuning $\Delta$. (\textbf{a}) $x$ and $y$ components of the RCP for a robust $X^{\pi}$ gate with a duration of $T_R=80$ ns. (\textbf{b}) Illustration of the two error curves ($\mathbf{r}^{\epsilon}$ and $\mathbf{r}^{\Delta}$) of the RCP in (\textbf{a}) for the amplitude and frequency noises. (\textbf{c}) QPT fidelities for the RCP (upper) and Gaussian (lower) pulse as a function of noise amplitudes $\epsilon$ and $\Delta$. Results from numerical simulation (right) and experimental data (left) are in good agreement. (\textbf{d}) The numerical result of the absolute difference of gate fidelity $|F_{RCP}-F_G|$ for the $X^{\pi}$ gate versus noise amplitude and decoherence times, where the RCP outperforms the Gaussian pulse in a broad region (above the red dashed line). Here the assumption is that $\epsilon=\Delta/\Omega_{\mathrm{max}}$ (noise) and $T_1=T_2$. The durations of the two gates ($T_R$, $T_G$) are 80 and 34 ns, respectively. }\label{Fig3}
    \end{figure}

    Following the same principles, one can readily design robust quantum gates for more realistic scenarios where noise and error appear in all three directions. Let us consider the following Hamiltonian $H=\frac{1}{2}(1+\epsilon)\Omega_x(t)\sigma_{x}+\frac{1}{2}(1+\epsilon)\Omega_y(t)\sigma_{y} + \frac{1}{2}\Delta\sigma_z$, where both a control amplitude noise $\epsilon$ and a frequency detuning $\Delta$ are present simultaneously. Using the aforementioned protocol, we construct the RCP for an $X^{\pi}$ gate, as shown in Fig.~\ref{Fig3}(a). Notice that the pulse now has both $x$ and $y$ components, which is necessary to suppress noises in all three directions. Experimental and simulated results of the gate fidelity extracted by QPT are plotted in Fig.~\ref{Fig3}(c). In the SI~\cite{SI}, we further discuss robust gates that can battle against errors due to noises with independent components in three directions, as well as the error induced by residual $ZZ$ coupling among qubits, which is notoriously harmful and difficult to handle in solid-state quantum systems. 

    The results shown in Figs.~\ref{Fig2} and~\ref{Fig3} clearly demonstrate the superior noise resilience of our robust gates. Besides this advantage, we also note that our RCPs are smooth pulses with relatively short durations compared to most other schemes of robust quantum gates. These features bring two important benefits. First, smooth pulses are more friendly for experimental implementation and also more likely to deliver reproducible performance. On the contrary, many existing protocols of robust gates adopt piecewise signals, often accompanied by abrupt jumps in parameters between different segments. Second, short duration always means less error due to decoherence. To further illustrate the practicality of RCPs, Fig.~\ref{Fig3}(d) numerically compares the performance of an RCP and a Gaussian $X^{\pi}$ gates in the presence of decoherence. It is clear that in a significant portion of the parameter space, RCPs outperform the Gaussian pulse by a large margin. This simulation is supported by the experimental results (see SI~\cite{SI} for a detailed comparison).      

    As discussed in the introduction, fidelity benchmarked by QPT alone does not faithfully herald the gate and circuit performance in realistic quantum circuits. For example, it is well known that noises of long correlations produce errors that accumulate coherently as a circuit progresses. The detrimental effect of such coherent errors becomes increasingly more prominent as the circuit depth grows. More severely, such coherent errors are difficult to perceive and correct. They thus pose great challenges for both QEC and NISQ applications~\cite{harper2023learning, iverson2020coherence, kueng2016comparing}. Common sources of such coherent errors widely observed in different platforms include non-Markovian noise of significant low-frequency components, long-term drift in systems, control imperfection, miscalibration, crosstalk, and unwanted qubit-qubit coupling of large characteristic times. In the following, we demonstrate how robust quantum gates, combined with other techniques, can help correct such temporally correlated coherent errors. For this purpose, we perform experiments on Clifford-based randomized benchmarking (RB)~\cite{knill2008randomized,magesan2012efficient,magesan2012characterizing} in the presence of a static frequency detuning, which represents an extreme case of temporal correlation. The qubit frequency is first calibrated as $\omega_{q0}$ with its neighboring couplers tuned far away. Then the frequency of one coupler is brought close to $\omega_{q0}$, so the actual frequency of the qubit becomes $\omega_q=\omega_{q0}-\Delta$. We determine $\Delta$ as a function of the bias applied to the coupler, then perform reference RB measurements at different values of $\Delta$. All gates in the RB sequences are designed assuming the frequency of the qubit being fixed at $\omega_{q0}$. Therefore, $\Delta$ appears as a static detuning. 
    
    \begin{figure}[t]
		\centering
		\includegraphics[width =0.48\textwidth]{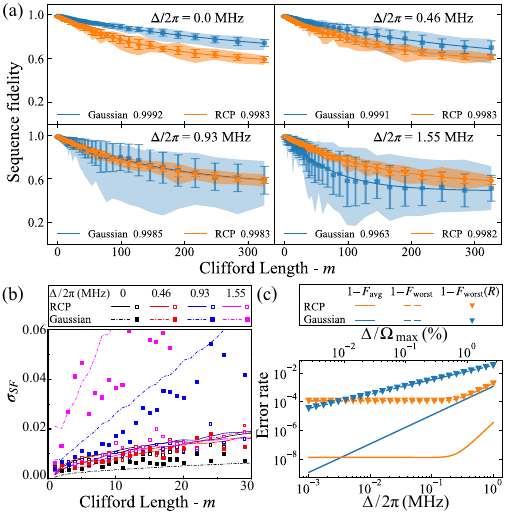}
		\caption{(\textbf{a}) Reference RB measurements are performed at different values of $\Delta$. Each measurement consists of 20 sequences of randomly chosen Clifford gates. Lines are fittings to the sequence fidelity by a formula $F = A p ^{m} +B$. The average fidelity per gate (lower legend in each panel) is given by $F_{\text{avg}} = 1-(1-p)/3.75$~\cite{magesan2012characterizing}. Error bars are the standard deviation and the shaded areas indicate the range of data. (\textbf{b}) Variance of sequence fidelity as a function of Clifford length. Lines are numerical simulations (see SI~\cite{SI}), and symbols are experimental results extracted from data in (\textbf{a}). (\textbf{c}) Numerical result of average and worst-case errors of the $X^{\pi}$ gate. Lines and symbols for the worst-case errors are calculated using the diamond norm solver and robustness measure respectively (see SI~\cite{SI})}.\label{Fig4}
    \end{figure}
    
    Figure~\ref{Fig4}(a) shows RB results obtained at different values of $\Delta$. At $\Delta=0$, the Gaussian case outperforms slightly due to a shorter average gate time, thus, less error resulting from decoherence. Also, the small oscillation of RCP's variance indicates some remaining correlated noises that probably come from the miscalibration of slightly more complex pulse shapes. As $\Delta$ increases, the performance of the robust gate sequences remains nearly unchanged, whereas the Gaussian sequences start to exhibit much-widened variances. Even though nominal high values of average fidelity per gate can still be extracted, they are definitely insufficient for evaluating the overall circuit performance. For example, the average error per gate for the Gaussian sequence is $\sim$0.08$\%$ for $\Delta/2\pi=0$ MHz, mainly caused by decoherence. It increases to $\sim$0.15$\%$ for $\Delta/2\pi=0.93$ MHz. However, the impact of such a seemingly small extra error resulting from the static detuning on the circuit performance is much more detrimental than errors due to decoherence. We also note that the maximum $\Delta/2\pi$ value studied here (1.55 MHz) corresponds to a relative frequency detuning of 0.023$\%$ ($\Delta_{\mathrm{max}}$/$\omega_{q0}$) and a relative noise magnitude of 4.1$\%$ ($\Delta_{\mathrm{max}}$/$\Omega_{\mathrm{max}}$), both being reasonably small in reality. But their impact cannot be neglected at all. 
 
    The built-in randomization in RB sequences is known to convert coherent errors induced by noises of long correlations into incoherent ones. Qualitatively speaking, this process is similar to a random walk in real space, where the distance of the walker drifting away from the starting point is determined by both the randomness in its moving pattern and the step size. In the current case, the step size corresponds to the average error accumulated in an individual gate, and the sequence fidelity variance resembles the distribution of a random walker's distance from its starting point. In Fig.~\ref{Fig4}(b), we plot the variance of sequence fidelity $\sigma_{SF}$ as a function of the Clifford length $m$. For the RB sequences using RCPs, the $\sigma_{SF}$-$m$ relation is nearly independent of the detuning $\Delta$, indicating that the dominant error accumulated in individual gates comes from decoherence, in sharp contrast to the Gaussian case. Of course, unlike a random walk in an Euclidean space, the accumulation of coherent errors in a quantum circuit proceeds in its Hilbert space and is described by the corresponding Pauli transfer matrices. The unique mathematical framework in the quantum case makes it difficult to carry out a quantitative analysis. More discussion on this issue is given in the SI~\cite{SI}.  
    
    Above, we show that combining randomization and robust quantum gates helps correct coherent errors induced by static detuning. It is reasonable to expect that coherent errors due to other physical processes with a quasi-static nature can also be corrected this way. On the other hand, coherent errors resulting from a generic noise with non-trivial time dependence but small correlation length, such as the control imperfection studied in Fig.~\ref{Fig3}, can be already handled by robust quantum gates without the need for randomization. Moving onto realistic quantum circuits where coherent errors of different correlation lengths exist and an intrinsic randomization as in RB is not available, one can then combine robust quantum gates with other techniques for mitigating coherent errors developed from a circuit perspective. One simple example is the so-called randomized compiling~\cite{wallman2016noise,urbanek2021mitigating,hashim2021randomized,gu2023noiseresilient}, in which effective randomization converts coherent errors into incoherent ones and can be routinely introduced into quantum circuits without sufficient built-in randomness.     

    The large variance observed in the RB sequences of Gaussian pulses has another severe implication: It means that the worst-case error of Gaussian gates is significantly larger in the presence of coherent errors~\cite{kueng2016comparing,ball2016effect,edmunds2020dynamically}. Within our theoretical framework, the worst-case error can be easily calculated from the robustness measure, that is, the error distance (see SI~\cite{SI}). Figure~\ref{Fig4}(c) plots the simulated average and worst-case errors of an $X^{\pi}$ gate for both Gaussian and RCP pulses. As the robustness of the worst-case error is concerned, the advantage of RCPs over the Gaussian pulses is even more prominent than only considering the average error. Since the worst-case error determines the fault tolerance of QEC codes~\cite{kueng2016comparing, aliferis2006quantum, aliferis2009fault}, our results are particularly beneficial for QEC applications.   
    
    Finally, we note that our protocol not only can help correct coherent errors as shown above, but it can also be used to realize robust control beyond single qubits more directly. In SI~\cite{SI}, we show that an iSWAP gate constructed using RCPs also exhibits excellent robustness against frequency fluctuations. Generalization for different two-qubit gate schemes in a variety of platforms is also straightforward.
    
    To conclude, we have demonstrated the effectiveness of robust single-qubit gates in correcting errors caused by spatially correlated noises, including unwanted $ZZ$ interaction, crosstalk, and temporally correlated noises, including frequency variation, pulse deformation in various directions and with diverse strengths. These noises are related to realistic physical processes observed in different platforms and cause coherent errors that are difficult to identify and correct. We have presented a good fit of experimental and theoretical results of process tomography. Our RB measurements have shown the suppression of the coherence of errors and reduction of the worst-case error. Our results offer a promising approach to combat the increasing challenges associated with correlated noises in integrated quantum chips. Additionally, when combined with other error mitigation methods, the techniques reported here help reduce such errors and may benefit both QEC and NISQ applications. Therefore, we propose an alternative path toward achieving high-quality large-scale quantum computing.

\bibliography{RobustGatesMain}
\bibliographystyle{apsrev4-2}

\end{document}


\title{Supplemental Information for ``Robust Quantum Gates against Correlated Noise in Integrated Quantum Chips"}
	
    \author{Kangyuan Yi}
    \thanks{K.-Y. Y. and Y.-J. H. contributed equally to this work.}
    \affiliation{Department of Physics, Southern University of Science and Technology, Shenzhen 518055, China}
    \author{Yong-Ju Hai}
    \thanks{K.-Y. Y. and Y.-J. H. contributed equally to this work.}
    \affiliation{Shenzhen Institute for Quantum Science and Engineering, Southern University of Science and Technology, Shenzhen 518055, China}
    \affiliation{International Quantum Academy (SIQA), Shenzhen 518048, China}
    \author{Kai Luo}
       \affiliation{Department of Physics, Southern University of Science and Technology, Shenzhen 518055, China}
	\affiliation{Shenzhen Institute for Quantum Science and Engineering, Southern University of Science and Technology, Shenzhen 518055, China}
     \author{Ji Chu}
      \affiliation{Shenzhen Institute for Quantum Science and Engineering, Southern University of Science and Technology, Shenzhen 518055, China}
    \affiliation{International Quantum Academy (SIQA), Shenzhen 518048, China}
    \author{Libo Zhang}
    \affiliation{Shenzhen Institute for Quantum Science and Engineering, Southern University of Science and Technology, Shenzhen 518055, China}
    \affiliation{International Quantum Academy (SIQA), Shenzhen 518048, China}
    \author{Yuxuan Zhou}
    \affiliation{Department of Physics, Southern University of Science and Technology, Shenzhen 518055, China}
	\affiliation{Shenzhen Institute for Quantum Science and Engineering, Southern University of Science and Technology, Shenzhen 518055, China}
    \author{Yao Song}
    \affiliation{Shenzhen Institute for Quantum Science and Engineering, Southern University of Science and Technology, Shenzhen 518055, China}
    \affiliation{International Quantum Academy (SIQA), Shenzhen 518048, China}
    \author{Song Liu}
    \affiliation{Shenzhen Institute for Quantum Science and Engineering, Southern University of Science and Technology, Shenzhen 518055, China}
    \affiliation{International Quantum Academy (SIQA), Shenzhen 518048, China}
    \affiliation{Guangdong Provincial Key Laboratory of Quantum Science and Engineering, Southern University of Science and Technology, Shenzhen, 518055, China}
    \author{Tongxing Yan}
	\email{yantx@sustech.edu.cn}
    \affiliation{Shenzhen Institute for Quantum Science and Engineering, Southern University of Science and Technology, Shenzhen 518055, China}
    \affiliation{International Quantum Academy (SIQA), Shenzhen 518048, China}
    \affiliation{Guangdong Provincial Key Laboratory of Quantum Science and Engineering, Southern University of Science and Technology, Shenzhen, 518055, China}
 
    \author{Xiu-Hao Deng}
    \email{dengxh@sustech.edu.cn}
    \affiliation{Shenzhen Institute for Quantum Science and Engineering, Southern University of Science and Technology, Shenzhen 518055, China}
    \affiliation{International Quantum Academy (SIQA), Shenzhen 518048, China}
    \affiliation{Guangdong Provincial Key Laboratory of Quantum Science and Engineering, Southern University of Science and Technology, Shenzhen, 518055, China}
    
    \author{Yuanzhen Chen}
    \email{chenyz@sustech.edu.cn}
	\affiliation{Department of Physics, Southern University of Science and Technology, Shenzhen 518055, China}
	\affiliation{Shenzhen Institute for Quantum Science and Engineering, Southern University of Science and Technology, Shenzhen 518055, China}
    \affiliation{Guangdong Provincial Key Laboratory of Quantum Science and Engineering, Southern University of Science and Technology, Shenzhen, 518055, China}
    \author{Dapeng Yu}
   \affiliation{Department of Physics, Southern University of Science and Technology, Shenzhen 518055, China}
    \affiliation{Shenzhen Institute for Quantum Science and Engineering, Southern University of Science and Technology, Shenzhen 518055, China}
    \affiliation{International Quantum Academy (SIQA), Shenzhen 518048, China}
    \affiliation{Guangdong Provincial Key Laboratory of Quantum Science and Engineering, Southern University of Science and Technology, Shenzhen, 518055, China}
	
	\date{\today }
	\vskip 0.5cm
	\maketitle
	
\section{I. Theoretical Framework of Robust Quantum Control} 
    
    We start with the generic Hamiltonian of a qubit driven by a control field $H_{c}(t)$ and subjected to a noise $V(t)$
    \begin{equation}
    	H(t) = H_{0} + H_{c}(t) + V(t),
    \end{equation}
    where $H_{0}$ is the free-qubit Hamiltonian. The noise may couple to the qubit from all directions and can be generally written as $V(t) = \sum_{jk} \delta^j_k(t)\sigma_{k}$. Hereafter we reserve the upper index $j$ for numbering different noise sources (i.e., we consider the general case of having multiple noise sources) and $k$ for $x,y,z$. The noiseless evolution unitary is $U_{0}(t)=\mathcal{T}\exp \{-i\int_{0}^{t}(H_{0}+H_{c}(\tau))d\tau \}$ ($\mathcal{T}$ stands for time ordering). We define the error unitary as $U_e(t)=\mathcal{T} \exp \{-i \int_0^t d \tau H_I(\tau)\}$ where $H_I = U_0^{\dagger} V U_0$ is the noise in the interaction frame determined by $U_0$. The total evolution unitary can be decomposed as $U = U_0 U_e$ because the following Schr\"{o}dinger equation holds
    \begin{align}
    \frac{dU}{dt} & = \frac{dU_0U_e}{dt}  \notag \\
    &= -i(H_0 + H_c)U_0 U_e - i U_0 H_I U_e  \notag \\
    &= -i(H_0 + H_c)U_0 U_e - i U_0 (U_0^{\dagger} V U_0) U_e \notag \\
    &= -i (H_0 + H_c + V) U \notag \\
    &= -i H U.
    \end{align}
    
    To realize robust quantum gates, we aim to find control pulses that meet the following conditions: with a given gate duration $T$, the control Hamiltonian generates a target gate operation while eliminating the impact of noise, i.e., $U_0(T) = U_{\text{target}}$ and $U_e(T) = I$. 
    
    To calculate the error unitary $U_e(t)$, we carry out a perturbation expansion of its generator~\cite{blanes2009magnus, zeng2019geometric, hai2022universal}. Up to the leading order, one has $U_e(t) \approx  e^{-i \sum_j \Phi^j(t)}$, where $\Phi^j(t)$ is a matrix associated with the error generated by the $j$-th noise source. We further divide the noise into a time-independent amplitude $\epsilon^j_k$ and a possibly time-dependent profile $v^j_k$ as $\delta^j_k(t) = \epsilon^j_k v^j_k(t)$ ($k=x,y,z$), so the noise in the interaction frame can be written as $H_{I}(t) = \sum_{j,k} \epsilon^j_k v^j_k(t) \mathbf{T}_{k}(t)\cdot \hat{\boldsymbol{\sigma }}$, where $\mathbf{T}_{k}$ is a unit vector determined by $\mathbf{T}_{k}\cdot \hat{\boldsymbol{\sigma }} = U_0^{\dagger}\sigma_k U_0$. The aforementioned $\Phi^j(t)$ is then given by
    \begin{align}
    	\Phi^j(t) &= \sum_{k=x,y,z} \epsilon^j_k \int_{0}^{t} d\tau v^j_k(\tau) U_0^{\dagger}(\tau) \sigma_k  U_0(\tau)  \notag \\
    	&= \sum_{k=x,y,z} \epsilon^j_k \int_{0}^{t} d\tau v^j_k(\tau) \mathbf{T}_k(\tau) \cdot \hat{\boldsymbol{\sigma }} \notag \\
    	&= \mathbf{r}^j(t) \cdot \hat{\boldsymbol{\sigma }}.
    \end{align}
    
    For the $j$-th noise source, the corresponding vector $\mathbf{r}^j(t)$ traces out a three-dimensional curve with a velocity of $v^j(t)$. This curve describes the accumulated impact of the noise on the evolution unitary and is thus named as \textit{error curve}. The crucial geometric correspondence between the noisy quantum dynamics and the error curve provides a powerful approach to robust quantum control. For example, the \textit{error distance} $R^j(t) = ||\mathbf{r}^j(t)||$ measures the susceptibility of the quantum dynamics to the $j$-th noise source and is thus a natural metric for characterizing robustness of quantum control. Essentially, a robust control requires the error distances to vanish at the gate duration $T$, i.e., $R^j(T) = 0$. With this condition met, one has $\Phi^j(T)=0$ and thus $U_e(T) = I.$ 
    
    In the following, we discuss robust quantum control for two specific types of noise. For simplicity, we apply the rotating frame transformation corresponding to the qubit frequency: $U_r = e^{-i H_0 t}=e^{-i(\omega_0 t/2) \sigma_z}$. In this frame, the total Hamiltonian becomes $H = H_c + V$. We study the $x$-$y$ control
    \begin{align}
    H_c(t) = \frac{1}{2}\Omega(t) \cos \phi(t) \sigma_{x} + \frac{1}{2}\Omega(t) \sin \phi(t) \sigma_{y}
    \end{align}
    with amplitude $\Omega(t)$ and phase $\phi(t)$. Generally speaking, a control term is able to suppress noises that are non-commute to it, and the $x$-$y$ control can deal with noises in all three directions. Specifically, we discuss longitudinal noise (frequency noise) and transverse noise (control amplitude noise) in the following subsections.

    \subsection{1. Noise in $z$ direction (frequency noise)}
    
    First, we consider a frequency noise in the form of $V=\epsilon v(t) \sigma_z$. The accumulated error matrix is
    \begin{align}
    	\Phi(t) &= \epsilon \int_{0}^{t} d\tau v(\tau) U_0^{\dagger}(\tau) \sigma_z  U_0(\tau) 
    	= \epsilon \mathbf{r}(t) \cdot \hat{\boldsymbol{\sigma }} 
    \end{align}
    The $\mathbf{r}(t)$ traces out a 3D space curve with speed $v(t)$. 
    In order to better understand the noisy dynamics, we study the description of the error curve in the Frenet-Serret frame in terms of the tangent, normal and binormal vectors $\{\mathbf{T},\mathbf{N},\mathbf{B}\}$.
    
    The unit tangent vector of the curve is $\mathbf{T} = \frac{\dot{\mathbf{r}}}{v} = U_{0}^{\dagger }(t)\sigma_{z}U_{0}(t)$, and its time-derivative defines the normal vector $\mathbf{N}$ as
    \begin{eqnarray}
    	\dot{\mathbf{T}}(t)\cdot\hat{\sigma} &=& iU_{0}^{\dagger}(t)[H_{0}(t),\sigma_{z}]U_{0}(t)  \notag \\
    	&=&\Omega (t)U_{0}^{\dagger }(t)(-\sin \phi (t)\sigma _{x}+\cos \phi
    	(t)\sigma _{y})U_{0}(t)  \notag \\
    	&=&\Omega (t)\mathbf{N}\cdot \hat{\sigma}
    \end{eqnarray}%
    The binormal vector is then given by $\mathbf{B}=\mathbf{T}\times \mathbf{N}$, 
    \begin{equation}
    	\mathbf{B}(t)\cdot \hat{\sigma}=U_{0}^{\dagger }(t)(-\cos \phi (t)\sigma
    	_{x}-\sin \phi (t)\sigma _{y})U_{0}(t)
    \end{equation}%
    Its time derivative satisfies 
    \begin{eqnarray}
    	\dot{\mathbf{B}}(t)\cdot \hat{\sigma} &=&\dot{\phi}(t)U_{0}^{\dagger }(t)(\sin
    	\phi (t)\sigma _{x}-\cos \phi (t)\sigma _{y})U_{0}(t)  \notag \\
    	&=&-\dot{\phi}(t)\mathbf{N}\cdot \hat{\sigma}
    \end{eqnarray}
    
    The three unit vectors $\{\mathbf{T},\mathbf{N},\mathbf{B}\}$, as the tangent, normal, and binormal unit vectors of the error curve formed a Frenet frame and satisfy the Frenet-Serret equations 
    \begin{equation}
    	\left( 
    	\begin{array}{l}
    		\dot{\mathbf{T}} \\ 
    		\dot{\mathbf{N}} \\ 
    		\dot{\mathbf{B}}%
    	\end{array}%
    	\right) =\left( 
    	\begin{array}{ccc}
    		0 & \kappa v  & 0 \\ 
    		-\kappa v  & 0 & \tau v \\ 
    		0 & -\tau v  & 0%
    	\end{array}%
    	\right) \left( 
    	\begin{array}{l}
    		\mathbf{T} \\ 
    		\mathbf{N} \\ 
    		\mathbf{B}%
    	\end{array}%
    	\right),
    \end{equation}%
    where $\kappa$ and $\tau$ are curvature and torsion of the curve $\mathbf{r}$ respectively. 
    Note that $\dot{\mathbf{T}}\cdot \mathbf{N} = \kappa v $ and $\dot{\mathbf{B}} \cdot \mathbf{N}= -\tau v $, we have the following geometric correspondence
    \begin{align}
    	\kappa(t) v(t) &= \Omega (t) \\
    	\tau(t) v(t) &= \dot{\phi}(t)
    \end{align}
    that link the noise profile, error curve, and the control Hamiltonian. For quasi-static frequency noise with $v = 1$ discussed in the main test, the amplitude and phase of the control pulses are directly related to the curvature and torsion of the corresponding error curves.

    \subsection{2. Noise in the amplitude of control pulse}
    
    For this type of noise, we have the following noise Hamiltonian
    \begin{align}
    V = \frac{1}{2}\epsilon_x\Omega(t) \cos \phi(t) \sigma_{x} + \frac{1}{2}\epsilon_y\Omega(t) \sin \phi(t) \sigma_{y}.
    \end{align}
    Here and in the main text, we assume that the error amplitudes in both $x$ and $y$ directions are equal: $\epsilon_x = \epsilon_y = \epsilon$. This is a reasonable assumption for the platform of superconducting quantum computation. The error matrix $\Phi(t)$ is given by
    \begin{align}
    	\Phi(t) &= \epsilon \int_{0}^{t} d\tau v(\tau) U_0^{\dagger}(\tau) (\cos\phi(t)\sigma_{x} + \sin\phi(t)\sigma_{y} ) U_0(\tau) 
    	= \epsilon \mathbf{r}(t) \cdot \hat{\boldsymbol{\sigma }},
    \end{align}
    where $v(t) = \frac{1}{2}\Omega(t)$. The Frenet vectors defining the error curve are
    \begin{align}
    	\mathbf{T}\cdot \hat{\sigma} &= U_{0}^{\dagger}(t)(\cos\phi(t)\sigma_{x}+\sin\phi\sigma_{y})U_{0}(t) \\ 
    	\mathbf{N}\cdot \hat{\sigma} &= - U_{0}^{\dagger}(t)(-\sin\phi(t)\sigma_{x}+\cos\phi\sigma_{y})U_{0}(t) \\ 
    	\mathbf{B}\cdot \hat{\sigma} &= U_{0}^{\dagger }(t)\sigma_{z}U_{0}(t)%
    \end{align}
    Therefore we arrive at a similar geometric correspondence
    \begin{align}
    	\kappa(t) v(t) &= \dot{\phi}(t) \\
    	\tau(t) v(t) &= - \Omega (t).
    \end{align}
    
    With the above correspondence, one can study the noise resilience of various control pulses via the corresponding error curves. It is also promising to construct robust control pulses using analytical and numerical methods to combat the noise by enforcing the explicit robustness condition (closeness of the error curve), as illustrated in the next section.

\section{II. Construction of Robust Gate}
    
    \subsection{1. Robust control pulses}
    
    All robust control pulses (RCPs) used in this work are obtained using an analytical-numerical pulse construction protocol established in \cite{hai2022universal}. We briefly summarize the protocol and provide information on the RCPs used in this work in this section. 
    
    To construct RCPs, we apply a GRAPE-like auto-differentiation gradient update for the pulse ansatz parameterized by its time-domain amplitude in order to minimize the cost function
    \begin{equation}
    C=( 1 - F(T) ) + \sum_j R^j(T),
    \end{equation}
    where $F$ is the fidelity evaluated between the evolution unitary generated by the pulse ansatz and the target gate defined in~\cite{pedersen2007fidelity}, $R^j$ is the error distance of the $j$-th targeted noise source and multiple noises can be considered by summing over $j$, $T$ is the gate time. With the cost function being reduced to near zero $C\approx 0$, we have $F(T) \approx 1$ and $R^j(T) \approx 0$ and the resulting control pulses satisfy the robust control conditions $U_0(T) = U_{\text{target}}$ and $U_e(T) = I$. Note that the error distance is a natural robustness measure that enables the construction of RCPs without accounting for the specific value of noise strength, which is different from the conventional numerical approaches of pulse optimization.
    
    Accompanied by the gradient update, we implement the following Fourier series (modified by a sine function) as an analytical constraint to enforce the function form of the pulse ansatz
    \begin{equation}
    	\Omega(t) = \sin(\frac{\pi t}{T})(a_{0}+\sum_{n=1}^{N}a_{n}\cos(\frac{2\pi n}{T}t + \phi_{n}) ) , 
    \end{equation}
    where $N$ is the number of Fourier components, which is fixed to $2$ for all robust gates demonstrated in the main text. $\{a_n, \phi_n\}$ are parameters to be determined. The RCPs used in our experiment are obtained by running the pulse construction algorithm until the cost function below a given minimum value $C < 10^{-7}$. We thus obtain the RCPs feature high fidelity and robustness with both error distance and best average infidelity at the level of $10^{-8}$ (as shown in Fig.4(c) in the main text).
    
    The parameters of the RCPs for all robust gates discussed in the main text are listed in Table~S\ref{tab:Pulse}. Here we fix the gate time to be $50$ ns. In practice, one can always rescale the pulses via $t \to \alpha t$ and $\Omega \to \Omega/\alpha$ to obtain RCPs for a different gate time. Such rescaling does not change the relative robustness measured in the dimensionless noise strength $\epsilon/\Omega_\text{m}$, where $\Omega_\text{m}$ is the maximal pulse amplitude.
    
    \begin{table*} %
    	\setlength{\tabcolsep}{20pt}
    	\renewcommand{\arraystretch}{1.3}
    	\begin{tabular}{llll}
    		\hline\hline
    		Gates & Pulses & $\{a_i\}$ (GHz) & $\{\phi_i\}$ \\ \hline
    		$X_{R}^{\pi}$ & $\Omega$ & $[0.01034,-0.25855,-0.03278]$ & $[-0.01523, -0.03790]$ \\ 
    		$X_{R}^{\pi/2}$ & $\Omega$ & $[0.34930,0.30764,0.00026]$ & $[-0.00305,-0.00609]$ \\
    		
    		\multirow{2}{3em}{$X_{\text{all}}^{\pi}$} & $\Omega_x$ & $[-0.13158,-0.65450,-0.42338]$ & $[0.00214,0.00734]$ \\
    		& $\Omega_y$ & $[-0.41686,-0.65453,-0.56110]$ & $[-0.00144,-0.00528]$ \\   
    		
    		\multirow{2}{3em}{$X_{\text{all, 2}}^{\pi}$} & $\Omega_x$ & $[0.00701, -0.23557, 0.03234, -0.24956]$ & $[0.00800, -0.60128, -0.02887]$ \\
    		& $\Omega_y$ & $[-0.32726, -0.12747, 0.16732, 0.06606]$ & $[0.03469, -0.07938, -0.09605]$ \\  \hline\hline
    	\end{tabular}
    	\caption{Parameters for RCPs with a gate time of $T=50$ ns. The first two gates ($X_{R}^{\pi}$, $X_{R}^{\pi/2}$) are used in the measurements in Fig.2 of the main text for correcting errors resulting from frequency noise. The $X_{\text{all}}^{\pi}$ gate is used in the measurements in Fig.3 of the main text for correcting errors due to both frequency and amplitude noise. The $X_{\text{all},2}^{\pi}$ gate is for correcting quasi-static noises with three independent components discussed in the following section.}
    	\label{tab:Pulse}
    \end{table*}

    \subsection{2. Independent Noise in Three Directions} 
    
\begin{figure}[htb]
    	\centering
    	\includegraphics[width=0.8\columnwidth]{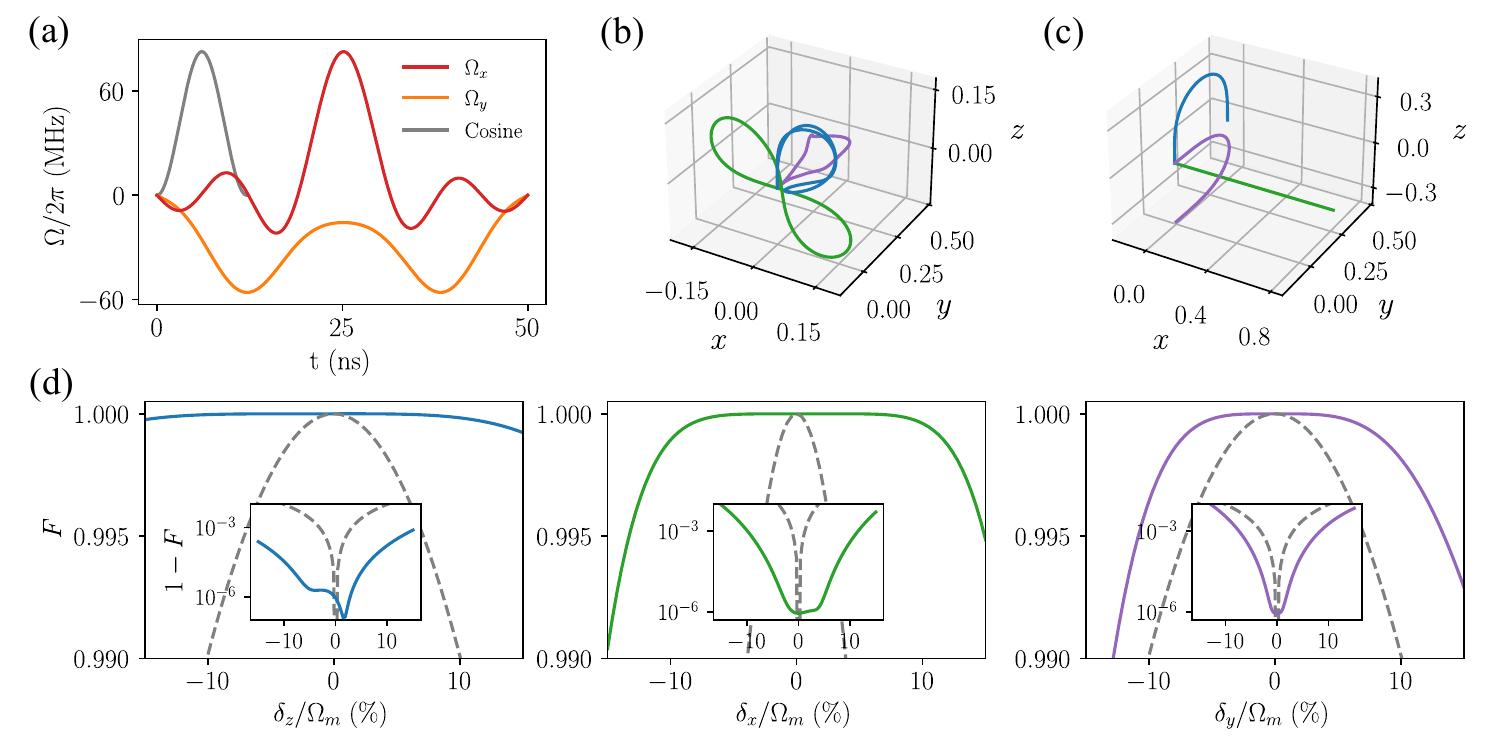}
    	\caption{ (a) An RCP (red and orange) and a cosine (grey) pulse for $X^\pi$ gate. (b)(c) Error curves of the RCP and cosine pulses for noises in $\{z,x,y\}$-direction (blue, green, purple). (d) Gate fidelity for the RCP (solid) and cosine (dashed grey) in the presence of static noise in $\{z,x,y\}$-direction. In each subplot, we vary the noise strength in one of the directions and fix the other two to zero.  }
    	\label{Fid_pulse_3noise}
    \end{figure}

    In the main text, we demonstrate a robust $X^\pi$ gate against both qubit frequency and control amplitude noise. Here we present another RCP for a robust $X^\pi$ gate against independent quasi-static noises in three directions. The total Hamiltonian is
    \begin{equation}
    H = \frac{1}{2}\Omega_x(t)\sigma_x + \frac{1}{2}\Omega_y(t)\sigma_y + \frac{1}{2}\delta_z\sigma_z + \frac{1}{2}\delta_x\sigma_x + \frac{1}{2}\delta_y\sigma_y,
    \end{equation}
    where we have control pulses in $x$ and $y$ directions with amplitude $\Omega_x$ and $\Omega_y$, respectively. $\{ \delta_z,\delta_x,\delta_y \}$ are the noise strength in $\{ z,x,y \}$ directions. This gate is denoted by $X_\text{all, 2}^{\pi}$ and the pulse parameters are listed in Table~S\ref{tab:Pulse}. We perform numerical simulation to compare it to a cosine pulse with a driving Hamiltonian $H_d = \frac{1}{2}\Omega_x \sigma_x$. The two control pulses and their error curves are illustrated in Fig.~S\ref{Fid_pulse_3noise}(a)-(c). Compared to the cosine pulse, the RCP has three closed error curves, indicating its error-robustness in three directions. The robustness is demonstrated by numerically calculating the gate fidelity versus different noise strengths, as shown in Fig.~S\ref{Fid_pulse_3noise}(d).
        
    Generally speaking, the higher the ratio of high frequency components are there in a pulse, the more Fourier components are needed for constructing the pulse. While in principle including more Fourier components helps reduce the discrepancy between the desired and the actual produced shape of a pulse, it also makes experimental implementation more complex and difficult. The optimal number of Fourier components used for constructing a pulse is often determined via try and error. For most pulses used in this work, it is found that two Fourier components are sufficient for producing both high fidelity and excellent robustness. On the other hand, the $X^{\pi}_{\text{all, 2}}$ pulse is designed to correct quasi-static noises in all three directions and contains relatively more high frequency components (compare, for example, the two solid orange lines in Fig. 2(a) and Fig. 3(a) of the main text). As a result, we have found that using three Fourier components can achieve a better performance in this case. 
    
    \subsection{3. Robust Single-qubit Gate against $ZZ$ Noise}
    
    \begin{figure}[htb]
    	\centering
    	\includegraphics[width=0.7\columnwidth]{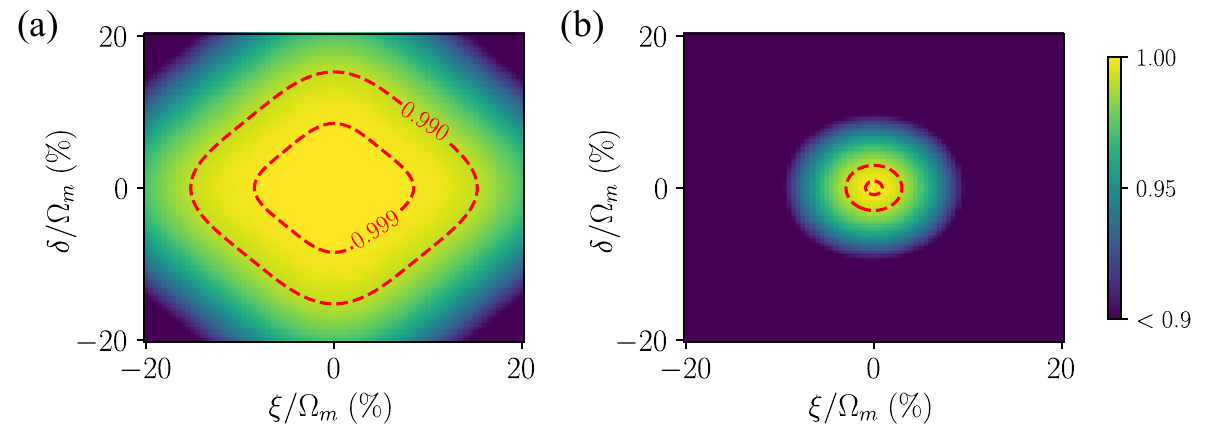}
    	\caption{ Gate fidelity of $X_R^{\pi/2}$ (a) and $X_C^{\pi/2}$ (b) pulses in the presence of static $ZZ$-coupling and single-qubit frequency noise. }
    	\label{Fid_zz_z}
    \end{figure}
    
    For single-qubit gates, besides the noise sources discussed in the main text and in the last subsection, our robust control protocol can also help correct errors resulting from unwanted coupling among qubits. Here we take the common $ZZ$ type of coupling as an example to illustrate. Residual $ZZ$ coupling among qubits is a major source of correlated noise in solid-state multi-qubit systems~\cite{krinner2020benchmarking, cai2021impact, xue2022quantum}. 
    
    Consider a single-qubit Hamiltonian in the rotating frame:
    $$
    H = \frac{1}{2}\Omega(t) \sigma_1^x + \frac{1}{2}\delta \sigma_1^z + \frac{1}{2} \xi \sigma_1^z \sigma_2^z,
    $$
    where both a frequency noise and a residual $ZZ$ coupling are present. Note that the control term does not commute with each of the noise terms. Up to the first order, we can separate the effect of different noises and the error unitary can be expressed as $U_e(t) \approx e^{-i(\Phi_z(t) + \Phi_{zz}(t))}$. The dynamics of the qubit corresponds to two error curves in the spaces spanned by $\{\sigma_1^{x},\sigma_1^{z},\sigma_1^{y}\}$ and $\{\sigma_1^{x}\otimes I,\sigma_1^{z}\sigma_2^{z},\sigma_1^{y}\sigma_2^{z}\}$. Fig.~S\ref{Fid_zz_z} plots the gate fidelity of two $X^{\pi/2}$ gates implemented by a RCP ($X_R^{\pi/2}$ in Table~S\ref{tab:Pulse}) and a cosine pulse. The RCP exhibits superior robustness than the cosine pulse counterpart against the two noises. We further note that the RCPs against frequency noise demonstrated in the main text also have the same robustness against the $ZZ$ noise. In practice, it is possible to combat correlated noises of different origins, such as unwanted interaction and interaction-induced crosstalk to further enhance the noise resilience of the quantum gates.

    \subsection{4. Robust Two-qubit Gate} 
    Within our framework, it is also possible to construct robust two-qubit gates. Take the iSWAP gate for example. Suppose two coupled qubits are subjected to quasi-static frequency noise $\delta_1$, $\delta_2$. Let $\Delta_{+}=(\delta_1+\delta_2)/2$ and $\Delta_{-}=(\delta_1-\delta_2)/2$, we have the rotating frame Hamiltonian for an iSWAP evolution
    \begin{equation} 
    \begin{aligned}
    	H & =-\frac{1}{2} \delta_1 \sigma_1^z-\frac{1}{2} \delta_2 \sigma_2^z+\frac{1}{2}g(t)\left(\sigma_1^x \sigma_2^x+\sigma_1^y \sigma_2^y\right)  \\
    	& =\left(\begin{array}{cccc}
    		- \frac{1}{2} \Delta_{+} & 0 & 0 & 0 \\
    		0 & - \frac{1}{2} \Delta_{-} & g(t) & 0 \\
    		0 & g(t) & \frac{1}{2} \Delta_{-} & 0 \\
    		0 & 0 & 0 & \frac{1}{2} \Delta_{+}
    	\end{array}\right) \\
    	& = - \frac{1}{2} \Delta_{+} (\sigma_1^z + \sigma_2^z) -
    	\frac{1}{2} \Delta_{-} (\sigma_1^z - \sigma_2^z) 
    	+ \frac{1}{2}g(t)\left(\sigma_1^x \sigma_2^x+\sigma_1^y \sigma_2^y\right).
    \end{aligned} \label{Eq_iSWAP_qubit}
    \end{equation}
    The $\Delta_{-}$ noise term can lead to a two-qubit over or under-rotation error. However, since this term does not commute with the drive term ($[\sigma_1^x \sigma_2^x+\sigma_1^y \sigma_2^y, \sigma_1^z - \sigma_2^z] \ne 0$), it can be suppressed by designing $g(t)$ into some RCP to achieve a robust control in the $\{|01\rangle,|10\rangle\}$ subspace. On the other hand, the $\Delta_{+}$ term commutes with the drive, but it only produces a single-qubit phase error, which can be more easily mitigated (e.g. by virtual $z$ correction) than the two-qubit error.
    
    \begin{figure}[htb]
    	\centering
    	\includegraphics[width=0.7\columnwidth]{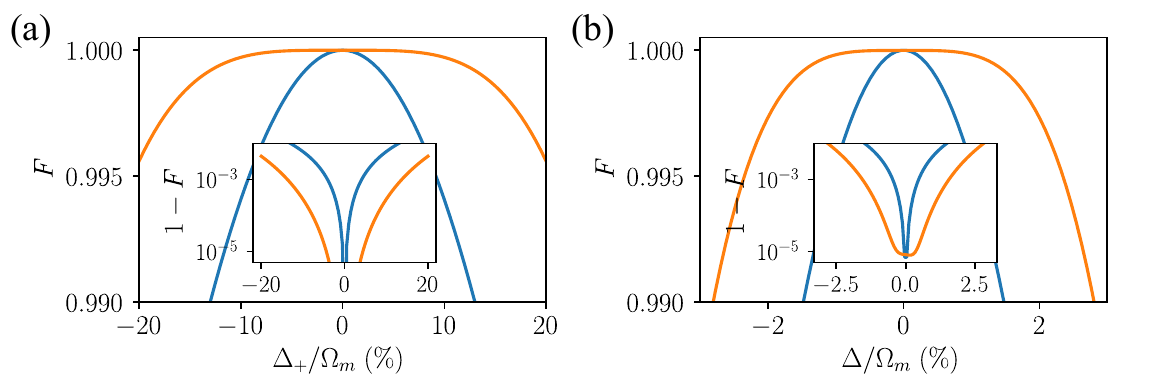}
    	\caption{ Robustness of iSWAP gate implemented by RCP (orange) and cosine pulse (blue) against qubit frequency noises. }
    	\label{Fid_iSWAP}
    \end{figure}
    
    Next, we apply the general idea discussed above to the specific case of two coupled superconducting transmon qubits with the following Hamiltonian
    \begin{equation}
    	H_{0}=\sum_{j=1,2}[\tilde{\omega} _{j}a_{j}^{\dagger }a_{j}+\frac{u_{j}}{2}%
    	a_{j}^{\dagger }a_{j}^{\dagger }a_{j}a_{j}]+g(a_{1}^{\dagger
    	}a_{2}+a_{1}a_{2}^{\dagger }),  \label{Eq_2Transmon}
    \end{equation}%
    The effective Hamiltonian in the two-qubit subspace up to the second perturbative order is
    \begin{equation}
    	H_{\text{eff}}=\sum_{j=1,2}\frac{1}{2} \omega_{j}\sigma_j^z + \frac{1}{2} \xi \sigma_1^z \sigma_2^z
    	+ \frac{1}{2}g\left(\sigma_1^x \sigma_2^x+\sigma_1^y \sigma_2^y\right) ,  \label{Eq_2Transmon_eff}
    \end{equation}%
    with $\Delta=\tilde{\omega}_1 - \tilde{\omega}_2$, $\omega_1 =-\tilde{\omega}_1-\frac{g^2}{\Delta}-\xi$, $\omega_2 =-\tilde{\omega}_2+\frac{g^2}{\Delta}-\xi$ and $ \xi =-\frac{g^2\left(u_1+u_2\right)}{\left(\Delta+u_1\right)\left(u_2-\Delta\right)}$. Note that there is an additional $ZZ$-coupling induced by higher levels of the transmons in the presence of the iSWAP interaction. The pulse for iSWAP itself cannot handle this $ZZ$ noise, but this noise can be compensated by appropriately tuning in the tunable coupler architecture~\cite{sung2021realization}. To better assess the robustness against the frequency noise, we omit the effect of this specific $ZZ$ noise in the numerical simulation.
    
    We numerically calculated the iSWAP dynamics for the qubit model (Eq.~(\ref{Eq_iSWAP_qubit})) and the transmon model (Eq.~(\ref{Eq_2Transmon})). The gate fidelity versus $\Delta_{-}$ and $\Delta$ noises respectively is shown in Fig.~S\ref{Fid_iSWAP}, where the iSWAP gates using an RCP and a cosine pulse are compared and the RCP exhibits great advantage. 
    
    Our robust control framework can be straightforwardly generalized for different two-qubit gate schemes in a variety of platforms for noise suppression, such as $\sqrt{\text{SWAP}}$ gate for quantum dot qubit, cross-resonance gate for transmon qubit and $XX$ gate for trapped ion qubit.

\section{III. Fidelity Measure}

    In this section, we discuss two common measures of fidelity of quantum gates, average fidelity and worst-case fidelity, in terms of the error distance defined in the previous section. The average fidelity of a quantum gate, by definition, only tells the average performance of that gate acting upon all possible input states. Of course, in certain circumstances, we would also like to know the worst possible performance of a quantum gate. The worst-case fidelity is defined for this purpose and often serves as a metric to determine fault-tolerance thresholds of QEC schemes. 
    
    Here we consider the case of a qubit suffering from a static detuning noise in $z$ direction: $V=\frac{1}{2}\delta \sigma_z$. The exact error unitary can be parameterized as $U_e=e^{-i \delta \sum_{k} \tilde{r}^{k} \sigma_k } = e^{-i \delta \tilde{\mathbf{r}}(t) \cdot \hat{\boldsymbol{\sigma }}} $. Different from the previous section, the $\tilde{\mathbf{r}}(t)$ defined here includes the contribution from all orders of the perturbative expansion. The total impact of the noise can be captured by the total error distance $R = \delta ||\tilde{\mathbf{r}}||^2$.
    
    The average fidelity is a standard measure for quantifying the performance of quantum gates and its benchmarking has become an experimental routine. Given an error channel $\mathcal{E}$, the average fidelity can be defined as
    \begin{equation} \label{Eq_avgFid}
    F_{\mathrm{avg}}(\mathcal{E})=\frac{\operatorname{Tr}[L(\mathcal{E})]+\operatorname{Tr}[\mathcal{E}(I)]}{d(d+1)},
    \end{equation}
    where $L(\mathcal{E})$ is the Liouville representation of the channel, $d$ is the dimension of the Hilbert space of the system. In the current case, $\mathcal{E} = U_e$ is a unitary error operator. In terms of the total error distance, the average gate fidelity can be expressed as
    \begin{equation} \label{Eq_avgFid_R}
    F_{\mathrm{avg}} = 1 - \frac{2}{3} \sin^2 R
    \end{equation}
    and the average error rate is $r = 1 - F_{\mathrm{avg}} = \frac{2}{3} \sin^2 R$.
    
    The worst-case error rate is defined as the diamond distance between the noise operator and the identity operator
    \begin{equation} \label{Eq_worstFid}
    D(\mathcal{E}-I)=\frac{1}{2}\|\mathcal{E}-I\|_{\diamond}=\frac{1}{2} \max _\rho\| I \otimes \mathcal{E}(\rho) - I \otimes I \|_1,
    \end{equation}
    where the maximum is taken over all density matrix of the system and $\|M\|_1=\operatorname{Tr} \sqrt{M^{\dagger} M}$. We emphasize that for certain scenarios relevant to quantum information processing, these two measures of error rate can differ by orders of magnitude~\cite{kueng2016comparing, sanders2015bounding}. According to~\cite{kueng2016comparing}, the diamond distance for unitary errors is bounded as follows
    \begin{equation} \label{Eq_worstFid_bound}
    \sqrt{\frac{d+1}{d}} \sqrt{r} \leq D \leq \sqrt{(d+1) d} \sqrt{r}.
    \end{equation}
    Therefore, for quantum gates that involve single or a few qubits, the worst-case error rate has the same order of magnitude as the square root of the average error rate. With small values of $r$, $D$ can thus be much larger than $r$. For the single qubit gates in this study, we take the lower bound of Eq.~(\ref{Eq_worstFid_bound}) for the worst-case error rate. In terms of the error distance, the worst-case fidelity then scales as
    \begin{equation} \label{Eq_worstFid_R}
    F_{\mathrm{worst}} \sim 1 - \sqrt{\frac{3}{2}} \sqrt{r} = 1 - | \sin R |.
    \end{equation}

    \begin{figure}[htb]
    	\centering
    	\includegraphics[width=0.65\columnwidth]{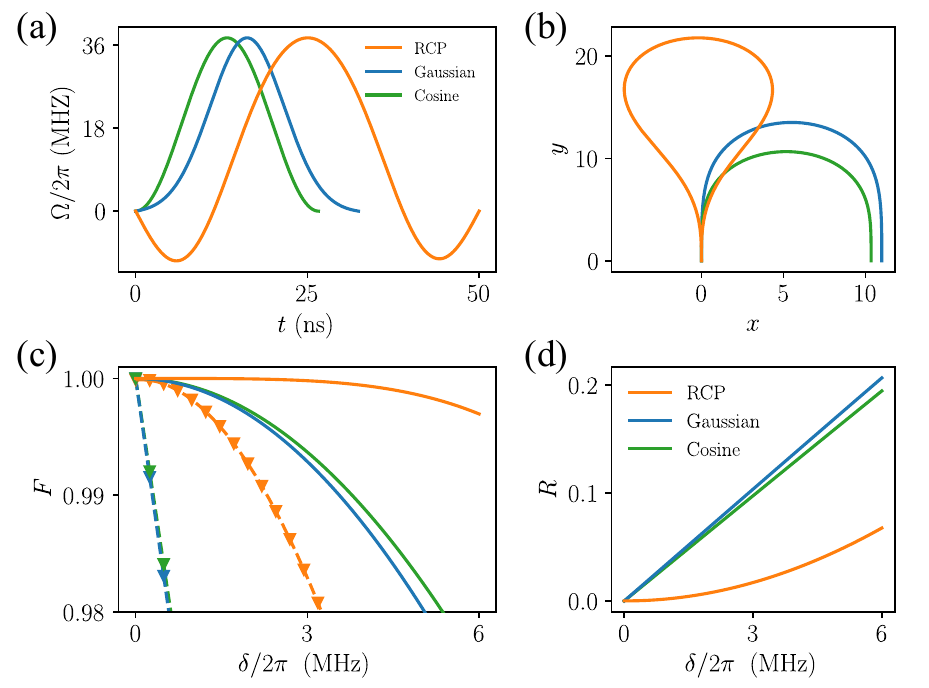}
    	\caption{ (a) Illustration of robust control pulse, Gaussian pulse and cosine pulse (orange, blue and green) for $X_{\pi}$ gate. (b) The error curves of the three control pulses. (c) Average gate fidelity (solid), worst-case fidelity (dashed) and the worst-case fidelity estimated by the total error distance (symbol) of the $X^{\pi}$ gate implemented by the three control pulses in the presence of qubit frequency noise. (d) Total error distance of the noisy quantum dynamics generated by the three control pulses as a function of the noise strength. }
    	\label{Fig_distance}
    \end{figure}
    
    We numerically verify the relation between the two fidelity measures and the total error distance. We study the error distance and fidelity of three types of control pulses for a single-qubit $X^{\pi}$ gate: the RCP, Gaussian pulse, and cosine pulse. The profile of the three pulses in the time domain and their error curves are illustrated in Fig.~S\ref{Fig_distance}(a)\&(b). Since the total error distance depends on the noise strength via
    \begin{equation} \label{Eq_total_R_Rz}
    	R = \delta R^z + O(\delta^2),
    \end{equation}
    the closeness of the error curve for the RCP indicates a vanishing error distance ($R^z = 0$) and first-order robustness, while the other two pulses have constant error distances $R^z = \text{const}$ and have no robustness.

    We numerically calculate the evolution unitary of the noisy quantum dynamics by solving the Schr\"{o}dinger equation and obtain the error unitary for each control pulse at different noise strengths. Then we calculate the corresponding total error distance by taking the logarithm of the corresponding error unitary and obtain the average gate fidelity through Eq.~(\ref{Eq_avgFid}) (or Eq.~(\ref{Eq_avgFid_R})). The total error distance of the three pulses is shown in Fig.~S\ref{Fig_distance}(d). The vanishing error distance and the vanishing total error distance (to the first order) of the RCP shown in Fig.~S\ref{Fig_distance}(b) and Fig.~S\ref{Fig_distance}(d) respectively explain the robustness of the average gate fidelity (solid lines in Fig.~S\ref{Fig_distance}(c)).
    
    We then calculate the worst-case fidelity via Eq.~(\ref{Eq_worstFid}) using the diamond norm semidefinite program solver in the QuTiP package~\cite{johansson2013qutip}. The results are shown as dashed lines in Fig.~S\ref{Fig_distance}(c). Obviously, the difference between the worst-case fidelities of the RCP and the other two non-robust pulses is even more significant compared to the difference between their average fidelities. According to Eq.~(\ref{Eq_worstFid_R}), the worst-case fidelity has an approximately linear dependence on the total error distance in the small noise region as
    \begin{equation} \label{Eq_worstFid_R_linear}
    F_{\mathrm{worst}} \sim 1 - R.
    \end{equation}
    Therefore, the worst-case fidelity of the Gaussian and cosine pulses in the small noise region is dominated by the non-vanishing error distance (first order term in Eq.~(\ref{Eq_total_R_Rz})) and deviates almost linearly from the identity, while that of RCP exhibits robustness and decreases more slowly. In Fig.~S\ref{Fig_distance}(c), We also plot the worst-case fidelity estimated from
    the total error distance (using Eq.~(\ref{Eq_worstFid_R_linear})) as symbols. The excellent agreement between symbols and solid lines indicates that the total error distance serves as a simple and good estimation for the worst-case fidelity so that we do not have to go through the more complicated calculations using Eq.~(\ref{Eq_worstFid}). 
    
    Given the special role of the worst-case fidelity in determining the fault-tolerant threshold, the width $\delta_m$ of the noise amplitude region $[0,\delta_m]$ in which the worst-case fidelity is higher than some threshold value $f$ becomes another valuable measure of control robustness. We take $f=0.99$ and observe $\delta_m/2\pi \approx 0.3 $ MHz for Gaussian and cosine pulses, while the ideal worst-case fidelity of the RCP still exceeds $0.999$ and the worst-case error rate is about two orders of magnitude lower than that of Gaussian and cosine pulses at this noise amplitude. For RCP, we have $\delta_m/2\pi \approx 2.3$ MHz, which demonstrates its excellent noise-resilient property. Our results underline the possibility of achieving superior error resilience in quantum gates via robust control techniques, which is highly valuable for realizing fault-tolerant quantum computing.
    
\section{IV. Device and Experimental Setup}
\begin{figure}[!htb]
        \centering
        \includegraphics[width =0.48\textwidth]{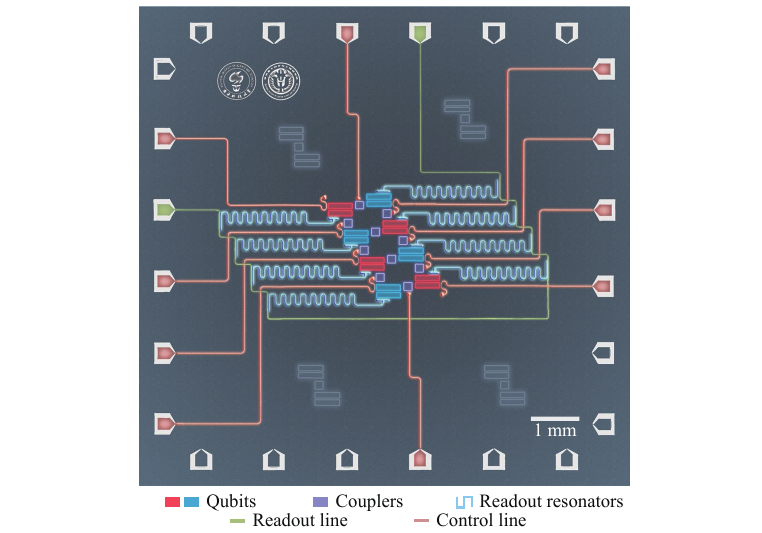}
        \caption{False-color micrograph of the experimental device. Processor design is the same as \cite{chu2023scalable,luo2023experimental}. \label{SM_8Qdevice}}
    \end{figure}

    We performed our experimental work on several different superconducting quantum processors. All data reported in this work were acquired on one processor that consists of 8 transmon qubits with fixed frequencies, arranged in a circle. Neighboring qubits are coupled to each other via couplers (also transmon qubits) with tunable frequencies. Figure S\ref{SM_8Qdevice} shows a false-color micrograph of this sample. Only one qubit and one coupler are used for the experiments in this work, whose parameters are summarized in Table S\ref{tab:qubitPara}.	
    
\begin{figure}[!htb]
        \centering
        \includegraphics[width =0.48\textwidth]{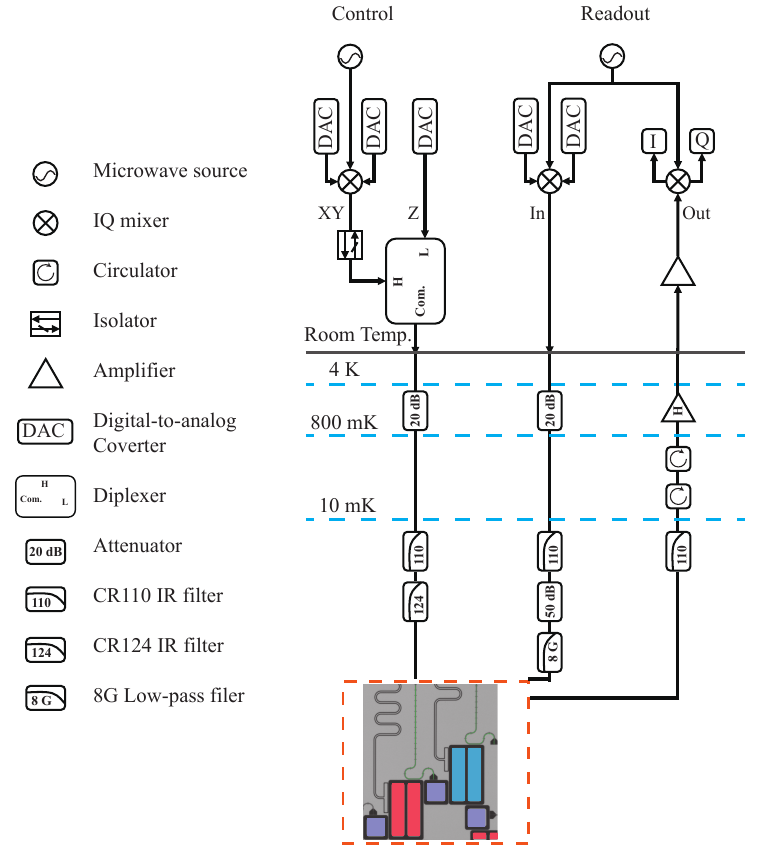}
        \caption{Schematics of wiring of the measurement setup.}\label{SM_wiring}
    \end{figure}

    The device is placed inside a dilution refrigerator with a base temperature of about 10 mK; details of the measurement circuitry of our experiment are shown in Fig.~S\ref{SM_wiring}.
    
    \begin{table}[ht]
        \begin{tabular}{ccc} 
            \hline  Parameters                         & Qubit   & Coupler \\
            \hline \hline
            Transition frequency, $\omega/2\pi$ (GHz)   & 6.734   & 7.7 \\
            \hline
            Anharmonicity, $\alpha/2\pi$ (MHz)          & -236    & -270 \\
            \hline
            Readout drive frequency, $f_{r}/2\pi$ (GHz)       & 4.980  & - \\
            \hline
            Qubit-Coupler coupling, $g_{qc}/2\pi$ (MHz) & \multicolumn{2}{c}{106} \\
            \hline
            Qubit energy decay time, $T_1$ ($\mu$s)                             & 19.8    & - \\
            \hline
            Qubit Ramsey decay time, $T_2^*$ ($\mu$s)                             & 22.5    & - \\
            \hline
            Qubit echo decay time, $T_2^E$ ($\mu$s)                             & 28.3    & - \\
            \hline
        \end{tabular}
        \caption{Parameters of the qubit and coupler used in the experiment.}
        \label{tab:qubitPara}
    \end{table}

\section{V. More Details of Experiments and Data Analysis}
    
    \subsection{1. Quantum process tomography}
    The single-qubit quantum gates studied in the main text are first characterized by the standard quantum process tomography (QPT). We prepare the qubit into 4 different initial states $\{|0\rangle,(|0\rangle - i|1\rangle)/\sqrt{2},(|0\rangle + |1\rangle)/\sqrt{2},|1\rangle\}$, then apply the gate operation to be benchmarked, followed by 3 rotations of
    $\{I,X^{\pi/2},Y^{\pi/2}\}$ to perform a standard quantum state tomography to reconstruct the final states. The experimental process  matrix $\chi_{\mathrm{exp}}$ is extracted using the formula $\rho_{f}=\sum_{m,n}\chi_{mn}E_{m}\rho_{i}E^{\dagger}_{n}$, which maps the ideal initial states $\rho_{i}$ into the experimentally obtained final states $\rho_{f}$\cite{nielsen_chuang_2010}. The orthogonal basis $E_n$ is taken as the Pauli matrices. We calculate the unattenuated gate fidelity using $F = |\text{Tr}(\chi_{\mathrm{exp}}\chi_{\mathrm{th}}^{\dagger})|/\sqrt{\text{Tr}(\chi_{\mathrm{exp}}\chi_{\mathrm{exp}}^{\dagger})\text{Tr}(\chi_{\mathrm{th}}\chi_{\mathrm{th}}^{\dagger})}$ to evaluate the QPT results\cite{zhang2012experimental,feng2013experimental,xu2018singleloop}. 
    
    \begin{figure}[!htb]
        \centering
        \includegraphics[width =0.7\textwidth]{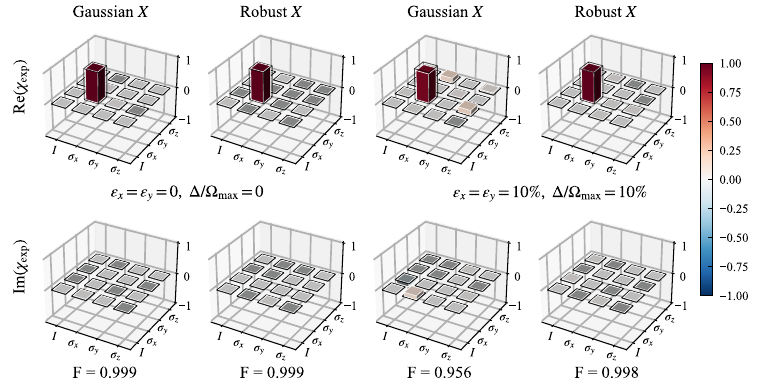}
        \caption{Process matrices for an $X^{\pi}$ gate using RCP and Gaussian pulses. The frameworks are theoretical values ($\chi_{\mathrm{th}}$) for an $X^{\pi}$ gate without any error. Bars are experimental results ($\chi_{\mathrm{exp}}$) extracted from QPT measurements performed at two different noise settings, both included in Fig.3(c) in the main text.}\label{SM_QPT}
    \end{figure}
    
    \begin{figure}[!htb]
        \centering
        \includegraphics[width =0.7\textwidth]{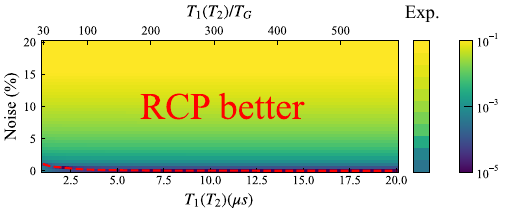}
        \caption{Left: Numerical simulation of the absolute difference of gate fidelity $|F_{RCP}-F_G|$ for an $X^{\pi}$ gate versus noise amplitude and decoherence times, where the RCP outperforms the Gaussian pulse in a broad region (above the red dashed line). Here the assumption is that $\epsilon=\Delta/\Omega_{\mathrm{max}}$ (noise) and $T_1=T_2$. The durations of the two gates ($T_R$, $T_G$) are 80 and 34 ns, respectively. The middle stripe is the experimental results of $|F_{RCP}-F_G|$ extracted from Fig.3(c) in the main text, with $T_1$ = 20 $\mu$s and $T_2$ = 25 $\mu$s, or $T_1$/$T_G$ = 588 and $T_2$/$T_G$ = 735. }\label{SM_multi}
    \end{figure}

    Figure S\ref{SM_QPT} shows the $\chi_{\mathrm{exp}}$ and corresponding QPT gate fidelity for several different cases studied in the main text. 
    
    Figure S\ref{SM_multi} shows a simulation similar to the one in Fig.3(d) of the main text but with a much larger range of the noise amplitude, and its comparison to the experimental results extracted from Fig.3(c) in the main text. The experimental results agree reasonably well with the simulation. 

    \subsection{2. Randomized benchmarking}
    
    In this study, the Clifford-based randomized benchmarking (RB) is used for two different motivations: to benchmark gate fidelity as a supplement to the QPT approach and to demonstrate the capability of our robust gates to correct temporally correlated errors. In the main text, we have thoroughly discussed the experimental work related to the second motivation using reference RB (RRB) measurements. There all quantum gates in the RB sequences are designed for a predetermined qubit frequency $\omega_{q0}$, and the actual qubit frequency $\omega_{q}$ while running these sequences is statically detuned from $\omega_{q0}$ by $\Delta$. Such a static detuning represents an extreme case of temporally correlated noise and produces a coherent error that is usually very difficult to perceive and correct in normal functional quantum circuits. On the other hand, the built-in randomization in the RB sequences converts such coherent errors into incoherent errors appearing across the collection of different sequences, therefore helping “visualize” the coherent error itself as well as the dynamical correction of the error by our robust gates. Of course, as explained in the main text, for quantum circuits without such a high level of randomness, one can still introduce effective randomization using, say, randomized compiling. 
    
    In this subsection, we fist provide results of additional RB measurements fulfilling the first motivation mentioned above. For this purpose, we performed both RRB and interleaved RB (IRB) measurements. We then discuss the impact of decoherence on the RB tests.

    \begin{figure}[!ht]
        \centering
        \includegraphics[width =0.7\textwidth]{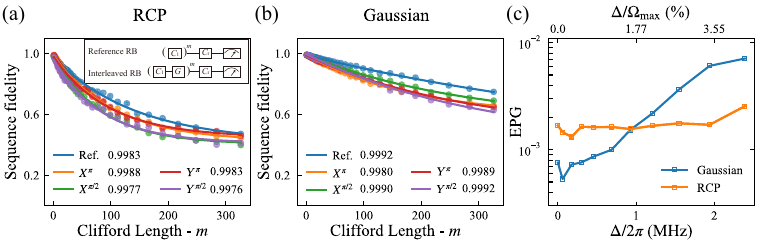}
        \caption{Experimental results of IRB benchmarking for four representative single-qubit gates using (a) RCPs and (b) Gaussian pulses. The frequency detuning $\Delta$ = 0 in these measurements. (c) Error rate per gate extracted from RRB measurements as a function of the frequency detuning $\Delta$. RRB data for $\Delta/2\pi$ = 0.0, 0.46, 0.93, and 1.55 MHz are shown in Fig.4 in the main text.}\label{SM_RB}
    \end{figure}
        
    Prior to each set of IRB measurements, we still need to first perform an RRB measurement to benchmark the average fidelity of all Clifford gates. This reference is then used in the following analysis of IRB measurements. The inset of Fig.~S\ref{SM_RB}(a) shows the circuits for both the RRB and IRB. For the RRB measurements, 20 sequences of randomly chosen Clifford gates are used and the average of their sequence fidelities (symbols) is used for extracting the average fidelity per gate. For this extraction, we fit the experimental data (i.e., the average of sequence fidelities, or the symbols) using $F_{seq} = A p^{m} + B$ as a function of Clifford length $m$ and obtain the decay rate $p$. For clarity, $p$ extracted in the RRB measurement is hereafter denoted as $p_{\text{ref}}$. Then the average fidelity per gate is calculated by $F_{\text{ref}} = 1 - (1-p_{\text{ref}})/3.75$. The IRB measurements use a similar protocol as the RRB measurement, except that we use $F_{\text{gate}} = 1 - (1-p_{\text{gate}}/p_{\text{ref}})/2$ to extract the fidelity of the interleaved gate. Here $p_{\text{gate}}$ indicates the decay rate obtained in IRB measurements. 
    
    Figure S\ref{SM_RB}(a)\&(b) show the experimental results of IRB measurements for four representative gates ($X^{\pi}$, $Y^{\pi}$, $X^{\pi/2}$, $Y^{\pi/2}$) using RCP and Gaussian pulses. In these measurements, the frequency detuning $\Delta$ is set to zero. The average fidelity per gate is obtained from the RRB measurements, and equals 0.9983 and 0.9992 for the RCP and Gaussian pulses, respectively. Since the average duration of RCP pulses is longer than that of Gaussian pulses, the RCP gates thus suffer more from decoherence. This fact accounts for the difference in the average fidelity per gate of the two cases. 
    
    To test the robustness of quantum gates against frequency detuning, we further run RRB measurements at different values of $\Delta$ and extract the error per gate as plotted in Fig.~S\ref{SM_RB}(c). The gates using RCP pulses exhibit a strong robustness against the frequency detuning. As a cost of such enhanced robustness, RCP gates have longer durations and thus under-perform the Gaussian gates at small values of $\Delta$ due to extra decoherence.
    
    \begin{figure}[!ht]
        \centering
        \includegraphics[width =0.7\textwidth]{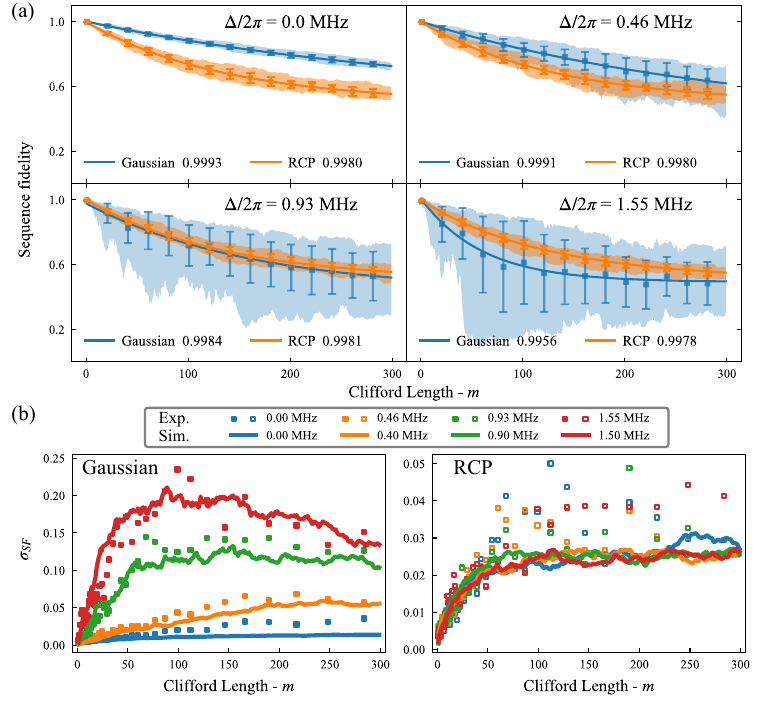}
        \caption{(a) Simulation of RRB experiments at different values of frequency detuning $\Delta$. The simulation uses identical conditions as those of the real RRB measurements reported in Fig.4 of the main text. For each simulation, 20 sequences of random Clifford gates are used. Lines are fittings to the average sequence fidelities of the 20 sequences. Error bars are standard deviations and shaded areas indicate the data range. Numbers in the insets are the extracted average fidelity per gate. For these simulations, we take $T_1$ = 20 $\mu$s and $T_2$ = 25 $\mu$s. (b) Variance of RRB measurements as a function of the Clifford length $m$ for different frequency detuning. Lines are results of RRB simulations composed of 100 sequences and symbols are extracted from the experimental results reported in Fig. 4 of the main text.} \label{SM_RBSimulation}
    \end{figure}
    
    Figure 4 in the main text and the related discussion gives a heuristic example that demonstrates the dynamical correction of temporally correlated coherent errors due to a static frequency detuning, by combining RB and our robust quantum gates. In the main text, we mentioned that the built-in randomization in RB changes the accumulation pattern of the coherent errors induced by quasi-static noises. An analogy can be made between the error accumulation in the RB case and a random walk in real space. However, there is an essential difference between the two scenarios. Unlike the Euclidean space in which a random walker wanders, the Hilbert space of a qubit where it evolves while accumulating errors is a compact space, therefore the ``distance" between the final and initial states of a qubit cannot increase unboundedly. Consequently, whereas the variance in the distribution of the final position of a random walker increases linearly and unboundedly over time, the variance in the sequence fidelity of an RB measurement behaves much more complicated. It does not grow unboundedly and exhibits non-monotonic trends over time. This fact makes a quantitative analysis of the variance in RB measurements rather difficult. For example, it may not be possible to extract useful information about the noise presented in an RB measurement. Nevertheless, we still tried to do some additional simulations and analysis of our RB measurements to achieve at least some qualitative understanding of the results.
     
    We first perform a simulation to see whether we can reproduce the main features of the RRB measurements reported in Fig.4 of the main text. Figure S\ref{SM_RBSimulation}(a) shows simulated RRB results under identical conditions as those of RRB experiments reported in Fig.4 of the main text. In this simulation, for each Clifford gate, we multiply its unitary matrix by two other matrices characterizing decoherence (both $T_1$ and $T_2$ processes), which can be obtained via the quantum channel of decoherence \cite{nielsen_chuang_2010}. The simulation produces the major features of the RRB measurements, including the nearly unchanged and much worsened performance of the RCP and Gaussian sequences, respectively, as the frequency detuning $\Delta$ increases. 
    
    Apparently, the variance of the sequence fidelity behaves very differently in the RCP and the Gaussian cases as $m$ increases for different values of frequency detuning. This fact indicates different speeds of error accumulation in the two cases. To understand the relation between the variance and the frequency detuning, we further perform simulations of RRB measurements composed of 100 sequences so that the variance changes more smoothly as $m$ changes. The results are plotted as lines in Fig.~S\ref{SM_RBSimulation}(b). For comparison, the experimental results of the real RRB measurements (i.e., those reported in Fig.4 of the main text) are also plotted as symbols here. Overall speaking, the simulated variance agrees with the extracted experimental results nicely. At small $m$, both Gaussian and RCP cases exhibit a nearly linear scaling of the variance versus $m$. For the RCP case, such linear scaling is almost independent of the frequency detuning, which means the average error accumulated in individual gates is insensitive to $\Delta$, whereas in the Gaussian case, it is very sensitive to the detuning. We also note that a larger variance means a lower worst-case fidelity, so it is obvious that the worst-case fidelity in the Gaussian case is also much lower than that in the RCP case. 

    As mentioned in the main text and at the beginning of this subsection, the error accumulation in the dynamics of a qubit is different from that in a random walk in real space. Therefore, a quantitative analysis of the behavior of the variance is rather difficult. Indeed, as $m$ increases, both the simulation above and our experimental results show complex trends for the variance. We have tried to establish some analytical model for the linear scaling behavior of the variance at small $m$ but have not achieved a conclusive result. Further studies are needed to gain a deeper understanding here. 
    
    \begin{figure}
        \centering
        \includegraphics[width =0.7\textwidth]{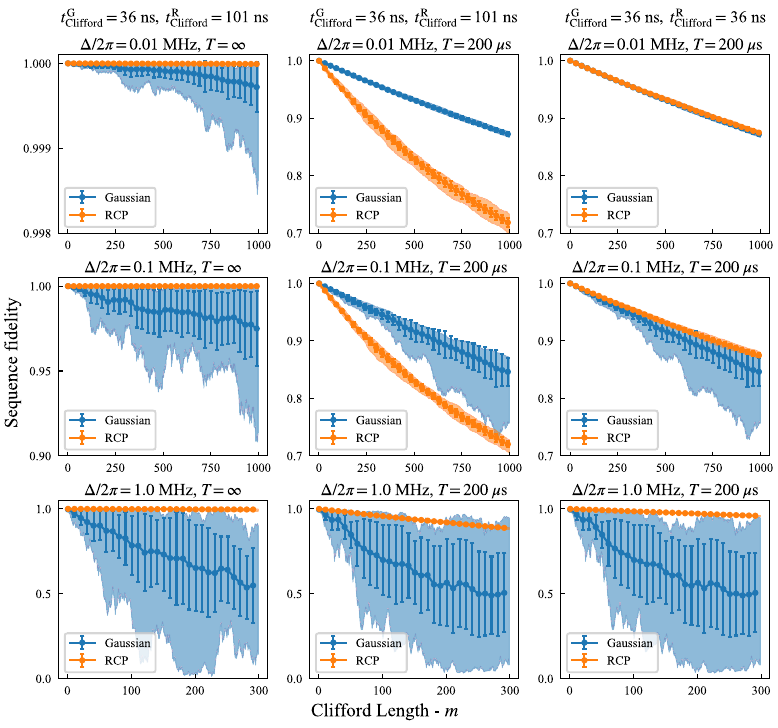}
        \caption{Simulation of RB experiments at different values of frequency detuning $\Delta$ and decoherence time $T$. The left and middle columns correspond to without decoherence and with decoherence times $T = T_1 = T_2 =$ 200 $\mu$s, respectively. The average duration per Clifford gate is around 101 and 36 ns for the RCP and Gaussian cases, respectively. In the right column, the decoherence time $T$ is the same as that in the middle column, but the average gate time per Clifford is set to be 36 ns for both RCP and Gaussian gates.} \label{SI_RBvsDecoherence}
    \end{figure}
    
    Lastly, we study the impact of decoherence on the performance of the RCP and Gaussian gates in RB tests. We first give an error budget analysis for the RB measurement in Fig. 4(a) of the main text, then present the results of additional numerical simulations. 
        
    In Fig. 4(a) of the main text, at $\Delta=0$, the average error per gate (obtained via RB) for the Gaussian and RCP cases are $7.6 \times 10^{-4}$ and $1.8 \times 10^{-3}$, respectively. Following the analysis given in Ref. \cite{omalley2015qubit,ding2023highfidelitya}, we estimate the incoherent error per gate due to decoherence to be $t_{\text{gate}}/T_{\text{error}}$, where $1/T_{\text{error}} = (1/T_1+1/T_{\phi})/3$ is calculated from qubit’s decoherence times. The average gate time in RB measurement for the RCP and Gaussian cases are 54 and 19.2 ns. Using the parameters of our device (see Table SII in the Supplemental Information), we obtain the incoherent error per gate for the RCP and Gaussian cases at $\Delta=0$ to be $\epsilon_{\text{incoherent, R}} = 1.3 \times 10^{-3}$ and $\epsilon_{\text{incoherent, G}} = 4.6 \times 10^{-4}$. Therefore, for the Gaussian case, 60.5\% of the total gate error is attributed to decoherence, while for the RCP case, 72.2\% of the total gate error is due to decoherence. 
    
    We further perform numerical simulations to study the impact of decoherence. The main results are summarized in Fig. S\ref{SI_RBvsDecoherence}. Here we consider a static frequency detuning for the qubit. 

    The left column shows the simulation results without decoherence. It is obvious that the RCP-RB shows nearly no decay for all three values of detuning (0.01, 0.1, and 1 MHz), indicating that the coherent errors due to such detuning are almost completely suppressed by the RCP gates, therefore its accumulation over a long RB sequence is still negligible. On the other hand, in the Gaussian-RB, the worst error after accumulating over 1000 gates is about 0.15\%, even though the average error per Gaussian gate is only $10^{-7}$ according to the simulation in Fig. 4(c) of the main text. For a moderate detuning of 0.1 MHz, the average error per Gaussian gate increases to about $10^{-5}$, which is still quite small. But the worst accumulated error in the RB after 1000 Clifford gates already amounts to about 10\%! These results vividly illustrate that the average gate error benchmarked by QPT is insufficient for evaluating the performance of Gaussian gates (in fact, any gate without built-in robustness) in a circuit where temporal-correlated errors exist. 

    The middle column shows the simulations with decoherence (here $T_1 = T_2 =$ 200 $\mu$s). The Gaussian case outperforms the RCP case at a detuning of 0.01 MHz. Since the RCP gates exhibit nearly no gate error at this detuning, the decay in the RCP-RB is exclusively due to decoherence. At a moderate detuning of 0.1 MHz, overall speaking the Gaussian-RB still looks better. However, we can already see the fast error accumulation in it as the circuit progresses. 

    For the two simulations discussed above, the average gate duration per Clifford gate is around 101 and 36 ns for the RCP and Gaussian cases, respectively. These are also actual gate durations used in our experiment. However, for qubits made of restrictive two levels or qubits with much larger anharmonicity (so leakage is far less an issue than in the transmon qubits), the duration of RCP gates can be designed to be on par with that of the Gaussian gates. To see how the two types of gates behave with similar durations, we have performed a simulation where the average duration per Clifford gate is set to be 36 ns for both RCP and Gaussian gates. The results are shown as the right column in the above figure. Now the RCP gates outperform the Gaussian gates in all cases. 

    Based on the above discussion, the apparent contradiction between Fig. 4(a) and Fig. 4(c) in the main text can be explained by the fact that the simulation in Fig. 4(c) did not take decoherence into account. 

    Again, we note that the inherent randomness of RB already greatly reduces the accumulation of coherent errors, which means the worst accumulated error in a general quantum circuit could be much worse than the RB case. This fact suggests that in general quantum circuits, the advantage of RCP over Gaussian may become even more prominent. 
    
    \subsection{3. Consideration on gate duration and suppression of leakage error}
    
\begin{figure}
        \centering
        \includegraphics[width =0.7\textwidth]{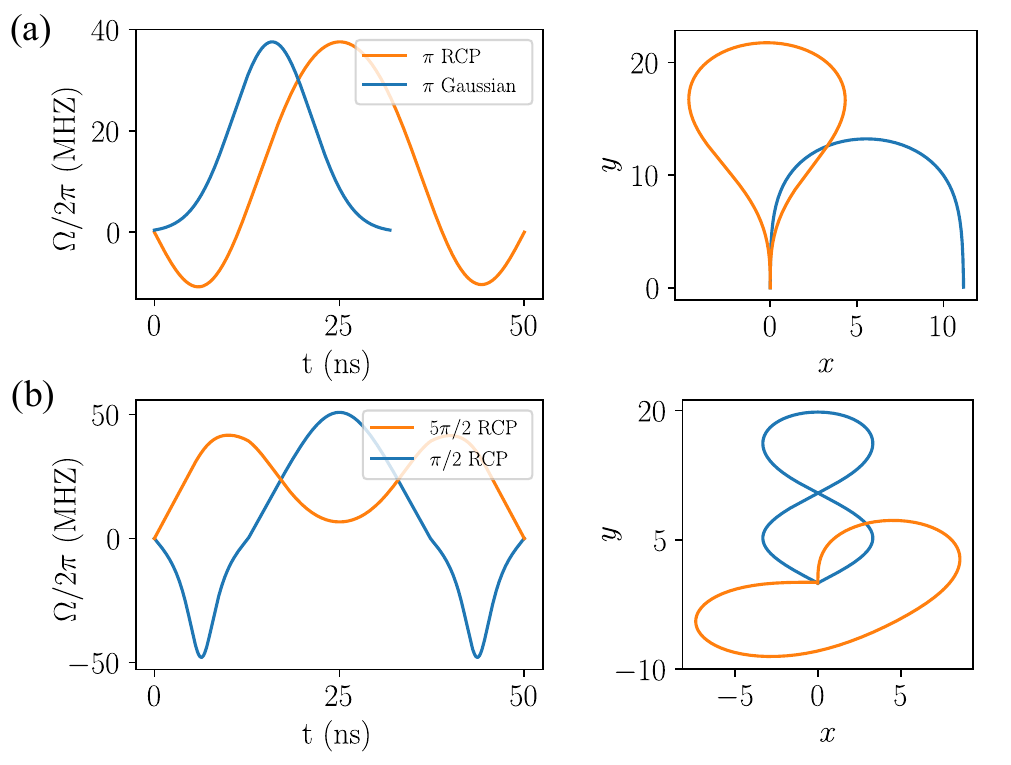}
        \caption{ Comparison of different control pulses. (a) RCP and Gaussian $\pi$ pulses for X-gate (left) and their error curves (right). (b) RCP $5\pi/2$ and $\pi/2$ pulses for X/2 gate (left) and their error curves (right).   } \label{SI_compare_pulse_pi}
    \end{figure}

    Let us first explain why an RCP-X gate is longer than a Gaussian-X gate. Figure \ref{SI_compare_pulse_pi}(a) compares the pulses of the two gates and their corresponding error curves (right) obtained via the geometric framework on which our work is based. Note here the two pulses are plotted showing their actual durations, which is different from Fig. 2 in the main text where the durations of both gates are normalized. 

    To facilitate a rigorous assessment of the robustness of various pulses against frequency detuning, it is imperative to standardize the maximal values of pulses, as pulse duration and bandwidth are intrinsically linked. Shorter pulses necessitate a higher maximal value to achieve the same unitary operation, and according to the Heisenberg uncertainty principle, this increased maximal value results in a broader bandwidth. This enhanced bandwidth allows for a more resilient response to frequency variations, ensuring the maintenance of high-fidelity gate operations despite potential deviations due to environmental or systemic imperfections. Hence, the robustness of a pulse against frequency detuning is directly proportional to its bandwidth, which is a function of the maximal pulse value. Consequently, setting identical maximal values for all pulses is a prerequisite for a fair comparison of their robustness to frequency detunings. 
    
    Within the geometric framework, imposing the first-order robustness against frequency detuning leads to an RCP-X pulse whose magnitude has both positive and negative sections, whereas the Gaussian-X pulse is always positive. Since the rotation angle ($\pi$ for both gates) is basically the integrated area under the pulses, it is obvious that the RCP-X pulse must be longer than the Gaussian-X pulse to accumulate the same angle. However, this fact is merely a consequence of the above constraint imposed on the signal magnitude. Without such a constraint, the RCP-X pulse can be much shortened by increasing its magnitude accordingly to keep the total rotation angle to be $\pi$. Theoretically, we can prove that such rescaling (reducing the duration while increasing the magnitude) does not compromise the robustness of RCP gates: their error curves stay closed upon such rescaling. 

    Of course, in realistic experiments, one must also consider other factors. For example, short pulses are more difficult to produce precisely. Moreover, short pulses may lead to more leakage errors for qubits with relatively small anharmonicity (e.g., transmon qubits). For qubits made of restrictive two-state systems or qubits with large anharmonicity so that leakage is far less a concern, we can largely shorten the duration of RCP pulses as long as they can be generated and transmitted in the experimental setup precisely enough. 
    
    Next, we explain why the RCP-X/2 pulse used in our work is even longer than the RCP-X pulse. An X/2 gate is normally realized by rotating the qubit around the $x$ axis by an angle of $\pi/2$. But in our work, it is implemented using an angle of $5\pi/2$. To see the reason for this choice, we need to examine the relevant RCP pulses in more detail. Figure \ref{SI_compare_pulse_pi}(b) shows two RCP-X/2 pulses of different rotating angles and their error curves. Note that the error curve corresponding to a $\pi/2$ angle consists of two closed loops. Remember that in our geometric framework, the rotation angle of a gate equals to the net angle swept by the tangent vector of the error curve as it traces along the curve. Geometrically it is impossible to construct an error curve whose tangent vector sweeps a net $\pi/2$ angle within one loop. The minimum number of loops is 2 for this purpose. 

    Alternatively, we can realize an RCP-X/2 gate by rotating the qubit around the $x$ axis by an angle of $5\pi/2$. For this choice, the error curve has only one loop. The corresponding pulse is more friendly for experimental implementation than the pulse for the case of a rotation angle of $\pi/2$, which contains sharp peaks that are more difficult to generate precisely and more likely to cause leakage error in transmon qubits.  

    To summarize, the actual choice of gate duration is the outcome of a weighted balance among several concerns: efficiency, technical complexity, and leakage error. In an ideal situation where the leakage is not a concern and accurate generation and transmission of pulses is guaranteed, there is no theoretical limit on the gate duration. 
        
    Unlike the ideal two-level model for qubits, transmons are nonlinear oscillators with a small anharmonicity. For the device reported in this work, the anharmonicity is -236 MHz (see Table S\ref{tab:qubitPara}). Fast gates with bandwidth comparable to this value thus suffer from leakage out of the computational space. Notice that our RCPs are compatible with the technique of Derivative Removal by Adiabatic Gate (DRAG)\cite{hai2022universal} widely used for suppressing leakage in multilevel qubits\cite{motzoi2009simple,gambetta2011analytic}. We employ DRAG for all the gates in our experiments. In addition, for a fair comparison, we implement RCPs and Gaussian pulses with different durations but the same maximum amplitude to match their bandwidths. In general, as a cost of enhanced robustness, the durations of the RCPs are longer compared to the Gaussian ones, which results in larger errors due to decoherence and relaxation. Nevertheless, our results indicate that the benefit of RCPs is overwhelming: quantum gates realized by these pulses show much improved robustness compared to the Gaussian gates. For qubits where leakage is much less a concern, one can reasonably expect that the application of our method can be more straightforward without the need for specially engineered leakage suppression; therefore, gate time can be significantly faster. 
    
\section{VI. Applicability of the RCP technique}

    In this last section, we discuss the applicability of the RCP technique in detail.

    \begin{figure}[!hbt]
        \centering
        \includegraphics[width =0.7\textwidth]{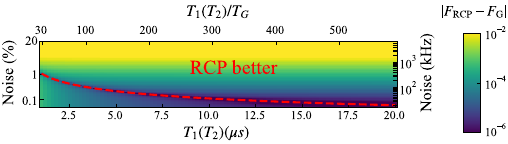}
        \caption{Figure S\ref{SM_multi} re-plotted with the noise axis in a logarithm scale so that the boundary (red dashed line) can be better seen.}\label{SI_simQPT}
    \end{figure}
    
    A careful examination of the experimental and simulation results (e.g., Fig. S\ref{SI_simQPT} and Fig. 4(c) in the main text) reveals the following fact: as far as the average fidelity or error rate is concerned (benchmarked via either QPT or RB), the difference between the RCP and Gaussian gates becomes appreciable only for detunings larger than a few hundred kHz. 

    Therefore a natural question to ask is what are the benefits of replacing Gaussian gates with RCP ones?
    
    We can answer such a question from two perspectives. First, relatively large frequency fluctuations or detunings (e.g., up to the MHz range) do exist in superconducting qubits and other platforms. For such large fluctuations, the advantage of RCP gates becomes obvious. Second, we emphasize that the advantage of RCP gates over Gaussian gates cannot be fully appreciated by merely comparing their average fidelity or error rate. Below, we elaborate further on these two points.

    \subsection{1. Large frequency fluctuation and detuning in superconducting quantum circuits and other platforms}

    Typical values of frequency fluctuations in transmon qubits are below 100 kHz\cite{burnett2019decoherence,vepsalainen2022improving}. However, this is only true when the qubits are in the so-called "normal" status. For superconducting qubits in realistic environment, we can identify at least three factors that can drive qubits out of such normal status. 
    
    First, qubits can couple strongly to defects such as two-level systems (TLS), which exist ubiquitously in solid-state devices. Frequency fluctuations of qubits due to TLS in the range of 100 kHz to several MHz have been reported\cite{klimov2018fluctuations,schlor2019correlating,lisenfeld2015observation,lisenfeld2019electric,omalley2015qubit}. While substantial progress has been made to the fabrication of superconducting quantum circuits in the last two decades, as manifested in a continuous improvement of decoherence times of superconducting qubits, TLS are still commonly observed by experimentalists. 
    
    There is currently no systematic solution to remove or mute TLS. Common practices of getting around the TLS problem include recalibrating qubits’ frequencies and choosing biasing points where the qubits’ frequencies are not close to that of any identified TLS. Both strategies have limitations. For the first one, as the superconducting quantum circuits scale up, the calibration procedure becomes ever increasingly complex. For those TLS having a characteristic timescale of a day or longer, it may be acceptable to recalibrate the circuit once a day or so to get around. However, this method certainly fails for TLS of shorter characteristic times. On the other hand, complex quantum circuits have very limited room for tuning qubits’ frequencies since it may lead to unwanted frequency crowding, let alone certain architectures that use qubits of fixed frequencies (see below). In this regard, the robustness against frequency detuning in gate operations offered by our RCPs can be useful for combating TLS in realistic solid-state quantum circuits. As long as the frequency detuning is not too large and too dynamic (i.e., being quasi-static within the time span of one experimental run, which has an upper bound set by the decoherence time on the order of 10-100 $\mu$s for today’s superconducting quantum chips), our method can help avoid frequent recalibration and rebiasing qubits, which becomes impractical for complex quantum circuits.  
    
    Second, in certain architectures of superconducting quantum circuits where qubits are designed to have fixed frequencies\cite{wei2022hamiltonian,krinner2020demonstration}, there may exist a strong $ZZ$ interaction between neighboring qubits leading to frequency detunings up to MHz for the coupled qubits. One solution to this problem is to design the processors so that the frequencies of neighboring qubits are rather detuned from each other, which helps reduce the frequency detuning. However, This practice often leads to problems such as frequency crowding among qubits and very long duration for two-qubit gates, etc\cite{gupta2024encoding,zhang2022highperformance}. Our RCP protocol offers a convenient alternative solution for such problems in addition to the existing efforts.   
    
    Third, control signals may also induce strong fluctuations in qubits' frequencies. For example, superconducting qubits are coupled to their readout resonators. If the resonator’s dissipation rate is small so that the average photon population in the resonator varies slowly, an effective quasi-static $ZZ$ interaction between the qubit and the resonator appears and has the form: $H_{int} = \chi \sigma_z a^{\dagger}a$, where $\sigma_z$ is the Pauli operator for the qubit and $a$ and $a^{\dagger}$ are annihilation and creation operators for the resonator modes. Typical value of $\chi$ is 0.2-5 MHz in superconducting circuits. This interaction also induces shifts in qubit’s frequency that depend on the average photon population in the resonator. Such frequency shifts may compromise the performance of QEC codes. When implementing QEC codes, one performs frequent measurements on the ancilla qubits, which also populates the readout resonators of the data qubits and thus induces frequency shifts to the latter via the mechanism discussed above. For fast readout required in QEC, $\chi$ is usually designed to be relatively high in the MHz range, which leads to frequency shifts of similar magnitude for the data qubits and results in dephasing from photon crosstalk between readout resonators\cite{googlequantumai2021exponential,szombati2020quantum}. To overcome this problem, various spin echo-based pulses were used to suppress the dephasing of the data qubits. However, due to exactly the same reason for shifted frequencies, such echo pulses introduce an additional bit-flip error in the course of suppressing the dephasing error. Using our RCP method may help correct the dephasing error without incurring the bit-flip error, achieving a better performance of QEC. 
    
    Above, we have discussed three factors that may cause large frequency fluctuations or detunings in superconducting qubits. We emphasize that this problem also exists in other platforms. For example, in silicon-based quantum dot systems, typical residual $ZZ$ interaction between qubits can reach a few hundred kHz to MHz\cite{xue2022quantum,watson2018programmable,connors2022chargenoise,madzik2020controllable}. Such a magnitude is comparable to the MHz frequency fluctuations in superconducting qubits discussed above since the magnitude of typical driving pulses in the Si-based quantum dot system is also at the level of a few tens MHz. However, the Si-based system differs from superconducting quantum circuits in two ways. First, the residual $ZZ$ interaction here cannot be completely turned off, whereas in superconducting circuits, one has various methods to do so. Second, Si-based qubits are perfect two-level qubits, unlike the transmon qubits which are inherently multiple-level systems. These two features make it even more beneficial and straightforward to implement RCP in Si-based qubits.  

    Lastly, in atomic quantum processors, the dephasing and pulse errors have become limiting factors for further improving fidelity\cite{bluvstein2022quantum,levine2018highfidelity,deleseleuc2018analysis,saffman2016quantum,jandura2023optimizing}. These errors largely come from the Doppler shifts, laser intensity and frequency fluctuations, noise in optical traps, and motional effects of atoms. Robust gate techniques are shown to help reduce the sensitivity of gate operations to these errors\cite{jandura2023optimizing}. Our RCP protocol can be useful here as well. 

    \subsection{2. Evaluating the advantage of RCP gates over Gaussian gates by the worst-case gate fidelity or error}

    One of the important messages we try to convey in this work is that the average gate fidelity or error obtained via QPT or RB is insufficient for evaluating gate performance in realistic quantum circuits. For example, QPT fidelity is obtained by averaging over different input states. But in realistic circuits, what matters is not the average fidelity, but the fidelity corresponding to specific input states. In addition, QPT only benchmarks the performance of individual gates and cannot reflect the effect of error accumulation in a circuit at all. On the other hand, while RB does consist of sequences of gates, its inherent randomness very effectively converts coherent errors into stochastic ones and thus drastically alters their accumulation rate. But randomness in realistic quantum circuits usually has a much lower level. These facts suggest that to fully analyze the advantage of the RCP gates, one must first go beyond the average fidelity and second consider error accumulation. 

    For the first aspect, one may look into the worst-case fidelity (or the worst-case error) in addition to the much more widely studied average fidelity. It is shown that for individual gates where coherent errors dominate\cite{kueng2016comparing}, the worst case error is proportional to $\sqrt{r}$, where $r$ is the average gate error obtained via, say, QPT. Therefore, for small $r$, the worst and average errors can differ by orders of magnitude. For example, an average error as small as 0.01\% results in a worst-case error of 1\%.  

    In Fig. 4(c) of the main text, we simulate the average gate error benchmarked via QPT (solid lines) and the worst-case error (dashed lines and symbols) for an RCP-X gate (orange) and a Gaussian-X gate (blue) in the presence of a frequency detuning. In the whole range of frequency detuning, the QPT error for both gates stays low. The maximum difference between the two QPT errors is about 0.1\% for a detuning of 1 MHz. However, when looking into the worst-case error, one reaches a different conclusion. In the whole range of detuning, the worst case error of the RCP gate stays below 0.2\%, whereas that of the Gaussian gate becomes larger than 1\% for a detuning above 100 kHz. The concept of worst-case error becomes particularly important when QEC applications are concerned since it can be directly compared to the threshold of fault-tolerance of QEC codes. A worst-case error larger than 1\% would jeopardize most known QEC codes. Therefore RCP may offer a significant benefit here. 
    
    To see the potential advantage of RCP gates over Gaussian gates in a realistic circuit, one must also take error accumulation into account. For this purpose, one refers to the concept of diamond distance $D$ discussed in sec. III above, which satisfies the following equation:

    \begin{equation} \label{Eq_worstFid_bound2}
    \sqrt{\frac{d+1}{d}} r \leq D \leq \sqrt{(d+1) d} \sqrt{r}.
    \end{equation}

    Note that this equation is different from Eq. \ref{Eq_worstFid_bound} where only unitary errors were considered. Here, the upper bound is saturated for the case where correlated errors dominate, and the lower bound is saturated for incoherent errors. Therefore, in the case of temporally correlated coherent errors (with the quasi-static frequency detuning being an extreme case), $D$ can be much larger than the average error $r$.   

    Unfortunately, the diamond distance is difficult to measure directly in experiments for a given circuit. In particular, RB is not a good benchmark to see the impact of accumulated temporal-correlated coherent errors, since the inherent randomness of RB is known to convert such errors into incoherent ones. As a result, in RB circuits, the worst-case error (or the diamond distance) saturates the lower bound discussed above. In other words, RB also only delivers an average benchmark of gate performance as QPT does. This fact explains why in our RB measurements, a significant advantage of RCP over Gaussian gates was observed only for relatively large frequency detunings. 

    \subsection{3. Applying the RCP technique in the presence of 1/$f$ noise}

    The quasi-static frequency detuning studied in this work is an extreme case of temporally-correlated noise. In reality, time-dependent low-frequency noises with a power law spectrum ($1/f$, for example) are dominant and have been widely observed in solid-state platforms. For example, transmon qubits with tunable frequency suffer from a flux noise which often exhibits a $1/f$ spectrum. In this subsection, we perform numerical simulation to study the performance of RCP gates when qubit frequency fluctuates with a $1/f$ spectrum. 
    
    \begin{figure}[htb]
    	\centering
    	\includegraphics[width=0.8\columnwidth]{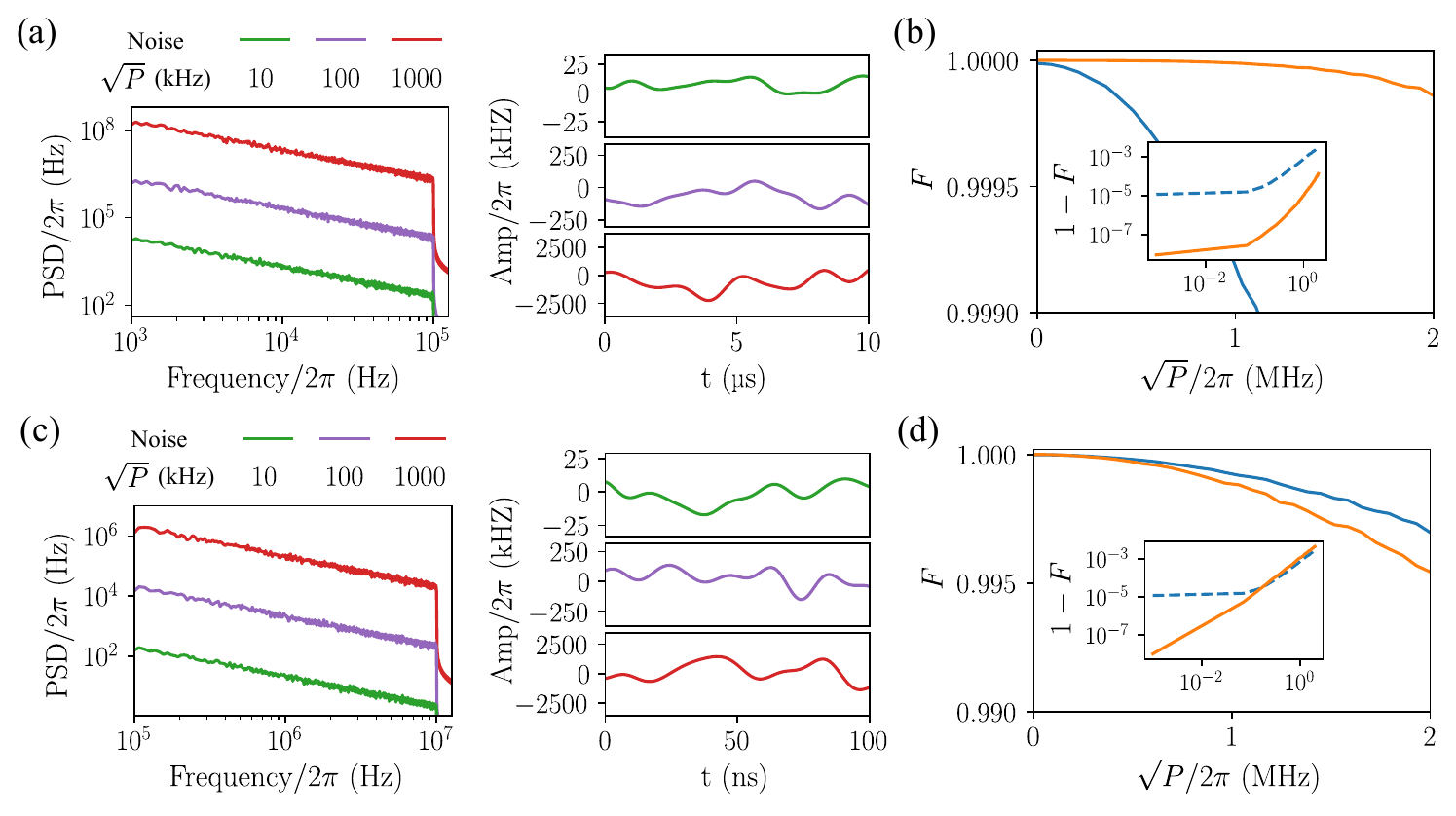}
    	\caption{ (a) $1/f$ noise within frequency range $1$ kHz to $100$ kHz. 
    		Left panel: Noise power spectrum density (PSD) with the power square root of $10$ kHz (green), $100$ kHz (purple), and $1000$ kHz (red). Right panel: typical samples of noise time series for the three different characteristic amplitudes (power square root). (b) Average gate fidelity as a function of noise power square root.
    	(c) $1/f$ noise within frequency range $0.1$ MHz to $10$ MHz. 
    	Left panel: Noise power spectrum density (PSD) with the power square root of $10$ kHz (green), $100$ kHz (purple), and $1000$ kHz (red). Right panel: typical samples of noise time series for the three different characteristic amplitudes (power square root). (d) Average gate fidelity as a function of noise power square root.  }
    	\label{Fig_1_f}
    \end{figure}

   The time series of $1/f$ noises can be generated by a combination of sine waves~\cite{done1992x, timmer1995generating}
      $$
         \delta(t) = \sum_i \sqrt{\frac{1}{f_i}} \sin(2\pi f_i t + \phi_i )
      $$
   where the frequencies $f_i$ are chosen within a desired range $f_i \in [f_{\text{min}},f_{\text{max}}]$ and $\phi_i$ are random phases in $[0,2\pi]$. The power spectrum density (PSD) of a noise can be calculated as a Fourier transform of its time series. 

   In our simulation, we consider $1/f$ noises within two representative frequency ranges of 1-100 kHz and 0.1-10 MHz and take 1000 evenly spaced frequency components for each noise series. For each frequency range, we generate $1/f$ noises of 30 different power levels ranging from 0 to 2 MHz. The choice of this range of power is in line with the typical experimentally measured values of long-term fluctuations in superconducting qubits' frequencies\cite{bylander2011noise,klimov2018fluctuations,schlor2019correlating,lisenfeld2015observation,lisenfeld2019electric,omalley2015qubit}. In Fig. S\ref{Fig_1_f}(a) we plot the spectra of three representative data sets of different powers and the corresponding time series. Each spectrum is an average of 50 independent noise samples, whereas each time series represents a single sample.  

   We then numerically calculate the average gate fidelity in the presence of the noises with different powers without decoherence. The fidelity for each power is obtained by averaging over 2000 noise realizations. Figure S\ref{Fig_1_f}(b)\&(d) compare the performance of an RCP-X gate and a Gaussian-X gate. For the case of a $1/f$ noise in the range of 1-100 kHz, the RCP-X gate significantly outperforms the Gaussian-X gate at all power levels. This result is not surprising since the characteristic time scale of the $1/f$ noise in this frequency range is still much larger than the gate duration, which means the noise can be still considered as ``quasi-static''. For the case of a $1/f$ noise in the range of 0.1-10 MHz, the RCP-X gate still outperforms its Gaussian counterpart at low noise power. As the noise becomes stronger, the Gaussian-X gate starts to have a very slight advantage. We note that in realistic solid-state devices for quantum computation, it is very common to observe fluctuations in qubits' frequencies with a $1/f$ spectrum in the range of 1-100 kHz reaching a power level of a few hundred kHz to MHz. Therefore our simulation results suggest that replacing Gaussian gates with the RCP gates does help improve circuit performance.

   In conclusion, our methodology remains efficacious for designing robust quantum gates, even in the presence of dynamic noise. Our theoretical framework is adaptable to time-varying noise profiles. By incorporating a model of fluctuating noises, a time-dependent function can supplant the static error vector, facilitating a dynamic traversal along the error curve with a variable velocity. 

\bibliography{SI}
\bibliographystyle{apsrev4-2} 